\documentclass[aps,prb,groupedaddress,showpacs,twocolumn]{revtex4}
\usepackage{graphicx}
\usepackage{amsmath}
\usepackage{amssymb}
\usepackage{bbm}
\usepackage{xspace}
\usepackage{color}

\newcommand{\matx}{\mat{x}}
\newcommand{\vcasc}{VCA$_{\text{SC}}$ }
\newcommand{\vcaom}{VCA$_{\Omega}$ }
\newcommand{\vcascNOBLANK}{VCA$_{\text{SC}}$ }
\newcommand{\vcaomNOBLANK}{VCA$_{\Omega}$ }
\newcommand{\GF}{\text{G}}
\newcommand{\gf}{\text{g}}
\newcommand{\TF}{\text{T}}
\newcommand{\GFBLANK}{\text{G }}

\newcommand{\TFBLANK}{\text{T }}

\definecolor{orange}{RGB}{252,77,6}
\definecolor{brown}{RGB}{200,127,50}
\definecolor{green1}{RGB}{00,100,00}
\definecolor{green2}{RGB}{00,150,00}
\definecolor{green3}{RGB}{00,200,00}
\definecolor{green4}{RGB}{00,250,00}

\newcommand{\fig}[1]{fig.\thinspace{}\ref{#1}}

\newcommand{\eq}[1]{eq.\thinspace{}(\ref{#1})}
\newcommand{\Eq}[1]{Eq.\thinspace{}(\ref{#1})}

\newcommand{\se}{sec.\@\xspace}

\newcommand{\app}{app.\@\xspace}

\newcommand{\etal}[0]{\textit{et al.}}
\newcommand{\tcite}[1]{ref.~\onlinecite{#1}}
\newcommand{\tcites}[1]{refs.~\onlinecite{#1}}

\newcommand{\Tr}[1]
{
\text{Tr}\,#1
}

\newcommand{\tr}[1]
{
\text{tr}\,#1 
}
\newcommand{\e}[1]{\times 10^{#1}}

\newcommand{\uu}{1\hspace{-3pt}1}

\def\bra#1{\mathinner{\langle{#1}|}}
\def\ket#1{\mathinner{|{#1}\rangle}}

\DeclareMathOperator{\sign}{sign}

\newcommand{\nag}{{\phantom{\dag}}}

\newcommand{\mat}[1]{\mathsf{#1}}

\begin{document}


\title{Variational cluster approach to the single impurity Anderson model}


\author{Martin Nuss}
\email[]{martin.nuss@student.tugraz.at}
\affiliation{Institute of Theoretical and Computational Physics, Graz University of Technology, 8010 Graz, Austria}
\author{Enrico Arrigoni}
\affiliation{Institute of Theoretical and Computational Physics, Graz University of Technology, 8010 Graz, Austria}
\author{Markus Aichhorn}
\affiliation{Institute of Theoretical and Computational Physics, Graz University of Technology, 8010 Graz, Austria}
\author{Wolfgang von der Linden}
\affiliation{Institute of Theoretical and Computational Physics, Graz University of Technology, 8010 Graz, Austria}


\date{\today}

\begin{abstract}
We study the single impurity Anderson model by means of cluster perturbation theory and the variational cluster approach (VCA). An expression for the VCA grand potential for a system in a non interacting bath is presented. Results for the single-particle dynamics in different parameter regimes are shown to be in good agreement with established renormalization group results. We aim at a broad and comprehensive overview of the capabilities and shortcomings of the methods. We address the question to what extent the elusive low energy properties of the model are reproducible within the framework of VCA. These are furthermore benchmarked against continuous time quantum Monte Carlo calculations. We also discuss results obtained by an alternative, i.e. self consistent formulation of VCA, which was introduced recently in the context of non equilibrium systems.
\end{abstract}


\maketitle
\section{\label{sec:introduction}Introduction}
In recent years both the applications for strongly correlated quantum impurity models and the number of successful approaches to harvest their physical results have grown enormously. Those models were introduced to describe the effects of magnetic transition metal impurities immersed in metallic hosts~\cite{anderson_localized_1961, friedel_electrical_1956}. Originally they were derived to capture remarkable physical properties like the resistance minimum~\cite{clogston_local_1962,kondo_resistance_1964} at a specific temperature scale $T_K$~\cite{hewson_kondo_1997} or the anomalous magnetic susceptibility and specific heat of such materials. Today a whole realm of applications for quantum impurity models has opened. They describe the physics of quantum dots and wires~\cite{meir_transport_1991, meir_low-temperature_1993, jauho_time-dependent_1994} as well as molecular electronics~\cite{hanson_spins_2007}. Applications range from nanoelectronics all the way to quantum information processing~\cite{loss_quantum_1998}. Their properties are essential for today’s technological applications in single electron transistors~\cite{kastner_kondo_2001} exhibiting the Coulomb blockade effect~\cite{sohn_mesoscopic_2009} or in devices dominated by RKKY interaction~\cite{ruderman_indirect_1954,kasuya_theory_1956,yosida_magnetic_1957}. The behavior of various magnetic phenomena and the fascinating branch of heavy fermion physics is described by strongly correlated quantum impurity models~\cite{steglich_superconductivity_1979,   coleman_heavy_2006}. Recent studies have shown that the remarkable material Graphene exhibits Kondo physics~\cite{chen_tunable_2011} which may be investigated theoretically by virtue of quantum impurity models. These models have further been applied to understand the adsorption of atoms onto surfaces~\cite{brenig_theory_1974,brako_slowly_1981,  langreth_derivation_1991}. In addition, they are of theoretical importance as solvable models of quantum field theories~\cite{wilson_renormalization_1975, affleck_quantum_2008}. A renewed interest in understanding and calculating dynamic quantities of these models was created with the advent of dynamical mean-field theory (DMFT)~\cite{georges_dynamical_1996, kotliar_electronic_2006, held_realistic_2006}. In the foundations of this theory quantum impurity models have to be solved as an auxiliary problem.\\
A wide range of methods and approximations have been suggested for the solution of quantum impurity models. They however prove to be a very delicate subject because standard perturbative approaches diverge~\cite{hewson_kondo_1997}. Prominent methods to gain physical conclusions include a self consistent perturbative expansion~\cite{yamada_perturbation_1975} and Bethe Ansatz techniques~\cite{bethe_zur_1931} for one dimensional problems. The low energy physics is very well described by numerical renormalization group (NRG)~\cite{bulla_numerical_2008} and in some limits also by functional renormalization group (FRG)~\cite{hedden_functional_2004, karrasch_finite-frequency_2008}, and density matrix renormalization group (DMRG)~\cite{nishimoto_density-matrix_2004,nishimoto_dynamical_2006,peters_spectral_2011}. There is a range of slave particle methods~\cite{florens_slave-rotor_2004, coleman_new_1984} available as well as methods based on Hubbard's X-operator technique~\cite{hubbard_electron_1963, lobo_atomic_2010} and calculations using variational wave functions~\cite{brenig_theory_1974}. Valuable physical insight has been gained by using equation of motion techniques applying different approximation schemes~\cite{haug_quantum_1996}. For moderate system sizes the Hirsch-Fye Quantum Monte Carlo (QMC)~\cite{hirsch_monte_1986} algorithm has proven to achieve good results. In the past years different approaches to continuous time QMC~\cite{gull_continuous-time_2010} have been applied very successfully to solve quantum impurity models especially in application with DMFT. In this context exact diagonalization (ED) methods have been explored to solve small systems~\cite{caffarel_exact_1994}.\\ 
As of today some limits of quantum impurity models are understood with great precision but there appear several gaps to be bridged. The low energy properties of these models are reproduced very well by renormalization group based approaches (i.e. NRG). These approaches in general have trouble to capture the high energy parts of the spectrum. The same may be said about QMC methods which if applicable yield dynamic quantities in imaginary time. The analytic continuation to the real energy axis is ill conditioned. Spectra obtained by, for example, the maximum-entropy method~\cite{gull_image_1978, skilling_classic_1989} have a large uncertainty for higher energies. Exact diagonalization methods, in principle, grant access to low as well as high energy parts of the spectrum at the same time. Due to the prohibitively large Hilbert space however only small systems (about ten to twenty sites) may be treated with this method, whose low-energy behavior is expected to deviate from the one of the infinite lattice significantly. Nevertheless, the advantage consists in the fact that the spectral properties may be determined directly on the real energy axis. Besides the issue of the low energy scale, also the flexibility to adapt to various impurity configurations and geometries is limited in many methods. NRG has been successfully applied only to the one and two impurity case so far. QMC approaches may suffer from the sign problem for more complex multiband models~\cite{gull_continuous-time_2010}. The region of large interaction strength is naturally difficult to treat in standard perturbative/diagrammatic approaches (i.e. diagrammatic perturbation theory or FRG).\\
In the present work we test cluster perturbation theory (CPT)~\cite{gros_cluster_1993,senechal_spectral_2000} and the variational cluster approach (VCA)~\cite{potthoff_variational_2003,potthoff_self-energy-functional_2003,potthoff_self-energy-functional_2003-1} on the single impurity Anderson model~\cite{anderson_localized_1961}. The great flexibility and versatility of CPT/VCA allows for obtaining approximate single-particle dynamic quantities and static expectation values in all parameter regions of any lattice impurity model with local interactions. However these many body cluster methods can not be expected to describe the low energy excitations as accurately as specifically tailored methods do. It is however interesting to see whether the correct low energy behavior may be reproduced at least to some extent. CPT as well as VCA bare several advantages~\cite{senechal_introduction_2008}: i) They yield spectra directly on the real axis and ii) also the high energy incoherent part of the dynamics becomes available. iii) They are applicable in all parameter regions and also at high interaction strengths. iv) They have the advantage of comparatively low computational cost for a required resolution. Our main goal in studying the well understood single impurity is to benchmark CPT/VCA for future application to the not so well understood case of multiband impurity models in various spatial geometries. This publication also sets the foundations for a future extension to non equilibrium problems.\\
The text is organized as follows. The single impurity Anderson model is introduced in \se~\ref{sec:model}. A short review on CPT and VCA in this context is given in \se~\ref{sec:method}. A self consistent formulation of VCA previously introduced in the context of non equilibrium problems~\cite{knap_nonequilibrium_2011} is presented in \se~\ref{sec:SC}. Some remarks about the choice of variational parameters are provided in \se~\ref{sec:Varpar}. In \se~\ref{sec:Omega} we discuss the grand potential $\Omega$ for infinite fermionic systems in relation with the VCA. Results for the single-particle dynamics of the SIAM are provided in \se~\ref{sec:Results}. In this section also the quality of the low energy Kondo physics is compared to benchmarking results from NRG, DMRG, CT-QMC, Hartree-Fock and Bethe Ansatz calculations. Finally, we summarize and conclude our findings in \se~\ref{sec:conclusion}.\\

\section{\label{sec:model}The single impurity Anderson model}
\begin{figure}
        \centering
        \includegraphics[width=0.48\textwidth]{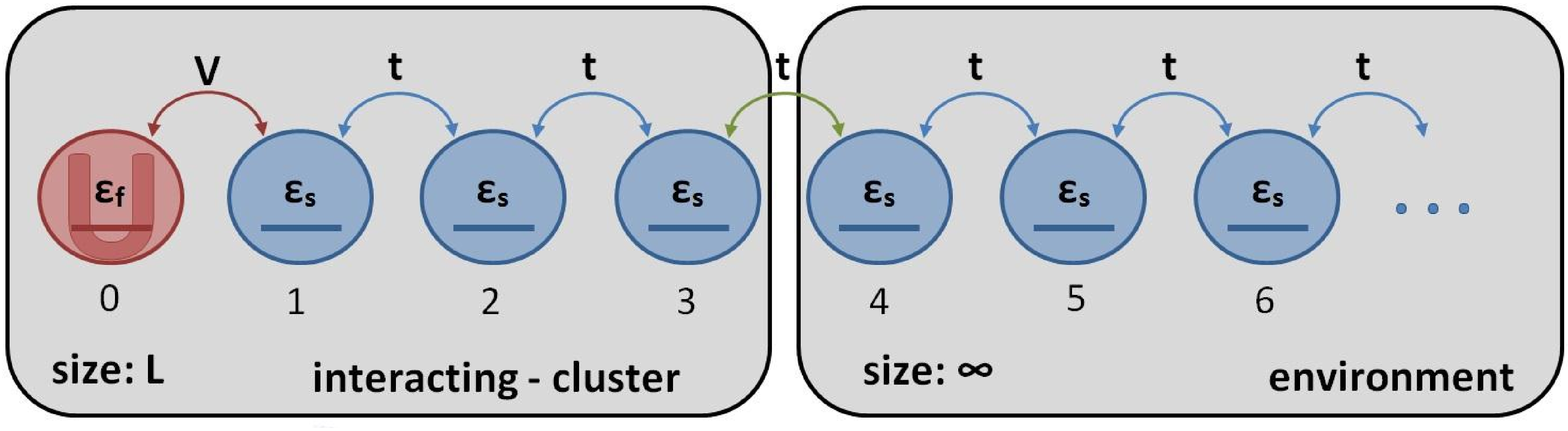}
        \caption{(Color online) 
Illustration of the single impurity Anderson model. The model consists of a semi-infinite chain of non interacting s-orbitals with nearest neighbor hopping $t$ and on-site energy $\epsilon_s$. The impurity f-orbital is subjected to a local on-site energy $\epsilon_f$ and local Coulomb interaction $U$ and is hybridized with one of the s-orbitals (here the one at the beginning of the chain) via a hybridization matrix element $V$. This maps the impurity f-orbital onto site $0$ and the impurity s-orbital onto site $1$ in this geometry, the rest of the conduction band s-electron orbitals are mapped onto sites $2$ to $\infty$. The semi-infinite non interacting chain is truncated at some site $L$. This decomposes the model into two clusters: an interacting cluster of variable size including the interacting impurity f-orbital and a semi-infinite chain of non interacting s-orbitals. In CPT/VCA these decomposed systems are coupled via a hopping element $t$.}
        \label{fig:SIAM}
\end{figure}

We consider the single impurity Anderson model (SIAM)~\cite{anderson_localized_1961} in real space, in one dimension 
\begin{align}
\hat{\mathcal{H}}_{\text{SIAM}} &= \hat{\mathcal{H}}_{\text{conduction}}  + \hat{\mathcal{H}}_{\text{impurity}}  + \hat{\mathcal{H}}_{\text{hybridization}} \;\mbox{.}
 \label{eq:HSIAM}
\end{align}
A tight binding band of non interacting s-electrons with nearest neighbor $\left\langle i,\,j \right\rangle$ hopping is described by
\begin{align}
\hat{\mathcal{H}}_{\text{conduction}}^{N_s} &= \epsilon_s \, \sum\limits_{i=1}^{N_s}\sum\limits_{\sigma}  \, c_{i\sigma}^\dagger \, c_{i\sigma}^\nag - t \, \sum\limits_{\left\langle i,\,j \right\rangle\,\sigma}  \, c_{i\sigma}^\dagger \, c_{j\sigma}^\nag \;\mbox{,}
 \label{eq:Hconduction}
\end{align}
where $\epsilon_s$ is the on-site energy of the particles, $t$ is the overlap integral between nearest neighbor orbitals and $i,\,j\in\{1,\ldots,N_s\}$ where $N_s$ is eventually taken to be infinity. The operators $c_{i\sigma}^\dagger$ and $c_{i\sigma}^\nag$, respectively, create and annihilate electrons in orbital $i$ with spin $\sigma$. The impurity Hamiltonian consists of a single f-orbital with local Coulomb repulsion $U$,
\begin{align}
\hat{\mathcal{H}}_{\text{impurity}} &= \epsilon_f \, \sum\limits_{\sigma}  \, f_{\sigma}^\dagger \, f_{\sigma}^\nag + U \, \hat{n}^{f}_{\uparrow} \, \hat{n}^{f}_{\downarrow} \;\mbox{,}
 \label{eq:Himpurity}
\end{align}
with $f_{\sigma}^\dagger$ creating an electron with spin $\sigma$ and on-site energy $\epsilon_f$ located at the impurity. The particle number operator is defined as $\hat{n}^{f}_{\sigma}=f_{\sigma}^\dagger\, f_{\sigma}^\nag$. Finally the coupling between a non interacting s-orbital and the impurity f-orbital is given by
\begin{align}
\hat{\mathcal{H}}_{\text{hybridization}} &= -V\, \sum\limits_{\sigma}  \, c_{1\sigma}^\dagger \, f_{\sigma}^\nag + f_{\sigma}^\dagger \, c_{1\sigma}^\nag \;\mbox{,}
 \label{eq:Hhybridization}
\end{align}
where $V$ is the hybridization matrix element between the s- and the f-orbital of the impurity atom (see \fig{fig:SIAM} for illustration).\\
We have set the chemical potential $\mu$ to the center ($\epsilon_s$) of the conduction electron density of states and choose $\mu = \epsilon_s = 0$. The resonance width $\Delta$ is defined as
\begin{align}
   \Delta \equiv \pi\,V^2\,\rho_s(0) &= \frac{V^2}{t}\;\mbox{.}
  \label{eq:Delta}
\end{align}
For the model defined in \eq{eq:Hconduction} the local density of states of the conduction electrons $\rho_s(0)$ is given by $\rho_s(0)=\frac{1}{\pi\,t}$. In the forthcoming discussion we refer to the particle-hole symmetric case when we furthermore set $\epsilon_f = -\frac{U}{2}$. All calculations are performed with $t=1$ and $V=0.3162$ which yields $\Delta=0.1$. All reported results, except for the CT-QMC data in \se~\ref{sec:CTQMC}, are for zero temperature.\\

\section{\label{sec:method}Cluster perturbation theory / variational cluster approach}
A handle on dynamic single-particle correlations and expectation values is given by the single-particle Green's function $\GF_{ij}^{\sigma\sigma'}(\omega)$ which we calculate within cluster perturbation theory (CPT)~\cite{gros_cluster_1993, senechal_spectral_2000} as well as the variational cluster approach (VCA)~\cite{potthoff_variational_2003}. CPT and VCA have been previously applied inter alia to the fermionic Hubbard model and VCA also with great success to bosonic systems~\cite{koller_variational_2006, arrigoni_extended_2011,knap_variational_2010} with broken symmetry phases. The groundwork of VCA lies within cluster perturbation theory which is a cluster extension of strong coupling perturbation theory, valid to first order in the inter-cluster hopping. The main result of CPT is that the  Green's function of the physical system $\GF$ (which we call full Green's function throughout this text) may be obtained by a Dyson-like equation in matrix form
\begin{align}
\GF^{-1} &= \gf^{-1} - \TF\; \mbox{.}
\label{eq:Dyson}
\end{align}
Here $\gf$ denotes the Green's function of a cluster which comes about by tiling the lattice of the physical system into smaller, numerically exactly solvable patches. This tiling is done by removing the hoppings between sites connecting such clusters. Therefore the matrix $\TF=\gf_0^{-1}-\GF_0^{-1}$ contains all single-particle terms connecting clusters (i.e. the inter cluster hopping which will be referred to as $\TF_{\text{inter}}$ below). The subscript~${}_0$ denotes the non interacting Green's function. To apply this approach to the SIAM we start by splitting the physical model under consideration (\eq{eq:HSIAM}) into appropriate pieces. Here we consider a cluster decomposition consisting of two parts. One part, consisting of a cluster of size $L$, which contains the interacting impurity f-orbital
\begin{align}
\hat{\mathcal{H}}_{\text{interacting}} &= \hat{\mathcal{H}}_{\text{conduction}}^{L-1} + \hat{\mathcal{H}}_{\text{impurity}}  + \hat{\mathcal{H}}_{\text{hybridization}} \; \mbox{,}
\label{eq:cluster}
\end{align}
and a second, infinitely large part, the environment, which contains the rest of the conduction band
\begin{align}
\hat{\mathcal{H}}_{\text{environment}} &= \hat{\mathcal{H}}_{\text{conduction}}^{\infty}\; \mbox{.}
\label{eq:environment}
\end{align}
The original Hamiltonian \eq{eq:HSIAM}, defined on the semi-infinite lattice, may now be rewritten as
\begin{align}
\hat{\mathcal{H}}_{\text{SIAM}} &= \hat{\mathcal{H}}_{\text{interacting}} +
\hat{\mathcal{H}}_{\text{environment}} + 
\TF_{\text{inter}}
\; \mbox{.}
\label{eq:Htiling}
\end{align}
Here $\TF_{\text{inter}}$ is the part of $\TF$ describing the hopping from the interacting cluster to environment ``cluster'', which is the only term not included in the two clusters. For the SIAM the two bare Green's functions $\gf_{\text{interacting}}$ and $\gf_{\text{env}}$ (, which correspond to $\hat{\mathcal{H}}_{\text{interacting}}$ (\eq{eq:cluster}) and $\hat{\mathcal{H}}_{\text{environment}}$ (\eq{eq:environment}), ) needed for \eq{eq:Dyson} may be evaluated separately. This is a bit different from the usual application of CPT to translationally invariant systems which normally leads to a single cluster having discrete spectra, embedded in a superlattice. Therefore the application of CPT to this problem does obviously not suffer from issues arising due to periodization prescriptions for the Green's function or self-energy~\cite{senechal_introduction_2008}. We are dealing with two fundamentally different clusters, where one has a discrete (interacting cluster \eq{eq:cluster}) and the other a continuous spectrum (environment \eq{eq:environment}). Due to the continuous spectrum of the environment a numerically favorable representation of the Green's function of the physical system $\GF$ in terms of the Lehmann representation (see for example \tcite{knap_spectralprop_2010}) is not possible. For evaluating quantities from the Green's function $\GF$ one therefore has to use a direct numerical integration, which works best on the Matsubara axis.\\
The cluster Green's function $\gf_{\text{interacting}}$ is determined by exact diagonalization of \eq{eq:cluster}. We apply the Lanczos algorithm~\cite{lanczos_iteration_1951} to find the ground state and a Band Lanczos method to obtain the Green's function. The Band Lanczos method is initialized with a set of all annihilation and creation operators
under consideration applied to the ground state. Thereby we obtain the so-called Q-matrices~\cite{zacher_evolution_2002} which are used to calculate the Green's function
\begin{align*}
 \gf_{\text{interacting},ij}^{\sigma\sigma'}(\omega) &= \sum\limits_\alpha \left( \sum\limits_\gamma Q_{i\gamma}^{\sigma} \frac{1}{\omega-\lambda_\gamma} Q_{j\gamma}^{\sigma'\dagger} \right)_{\alpha}\\
 Q_{i\gamma}^{\sigma\dagger} &= \begin{cases}\frac{1}{\sqrt{d}}<\gamma|\hat{c}_i^{\sigma\dagger}|\Psi_0>&\text{particle part}\\\frac{1}{\sqrt{d}}<\Psi_0|\hat{c}_i^{\sigma\dagger}|\gamma>&\text{hole part}\end{cases}\\ 
\lambda_{\gamma} &= \begin{cases}\omega_\gamma-\omega_0&\text{particle part}\\ \omega_0-\omega_\gamma&\text{hole part}\end{cases}\;\mbox{.}
\end{align*}
Essentially this is the Lehmann representation for zero temperature Green's functions. The sum over $\alpha$ denotes a sum over a possibly $d$-fold degenerate set of ground states. The sum over $\gamma$ is over a set of orthonormal basis-states having one particle more than the ground state (particle part) and one particle less than the ground state (hole part).\\
The Green's function of the environment $\gf_{\text{env}}$ is given analytically by the Green's function of a semi-infinite tight binding chain~\cite{economou_greens_2010}
\begin{align}
 \gf_{\text{env},i,j}(\omega) &= \upsilon_{0,i-j}(\omega) - \upsilon_{0,i+j}(\omega)   \label{eq:env}\\
\nonumber  \upsilon_{i,j}(\omega) &= \frac{-i\,\sign\left(\Im{\text{m}(\omega)}\right)}{\sqrt{4 |t|^2-(\omega-\epsilon_s)^2}}\,\Bigg( -\frac{\omega-\epsilon_s}{2|t|} \\
\nonumber  &+ i\,\sign\left(\Im{\text{m}(\omega)}\right)\,\sqrt{1-\left( \frac{\omega-\epsilon_s}{2|t|} \right)^2}\Bigg)^{|i-j|}\;\mbox{,}
\end{align}
where $\upsilon_{i,j}$ is the retarded / advanced Green's function of the infinite tight binding chain if the infinitesimal imaginary part ($0^+$) of $\omega$ is positive / negative.\\
VCA, the variational extension of CPT, is based on the self-energy functional approach (SFA)~\cite{potthoff_self-energy-functional_2003, potthoff_self-energy-functional_2003-1}.
In the SFA one considers the Legendre transformed Luttinger-Ward~\cite{luttinger_ground-state_1960} functional $F[\Sigma]$, which is a universal functional of the self-energy, i.e. it does not depend on  $\GF_0$. $F$ generates the Green's function, i.e.
\begin{align}
\beta\,\frac{\delta F[\Sigma]}{\delta \Sigma} &= -\GF[\Sigma]\;\mbox{,}
\label{eq:vca1}
\end{align}
where $\beta$ denotes the inverse temperature. Introducing the (non-universal) 
self-energy functional
\begin{align}
\Omega[\Sigma, \GF_0] &= F[\Sigma] - \text{Tr}\ln{ \left( \left(-\GF^{-1}_0 + \Sigma\right)\GF_\infty\right)} \,\mbox{,}
\label{eq:vca3}
\end{align}
(see \tcite{arrigoni_extended_2011} for a definition of $\GF_\infty$), one recovers
Dyson's equation at its  stationary point
\begin{align}
\beta\, \frac{\delta \Omega[\Sigma, \GF_0]}{\delta \Sigma} &= -\GF[\Sigma] + \left( \GF^{-1}_0 - \Sigma \right)^{-1} \stackrel{!}{=} 0\,\mbox{.}
\label{eq:vca2}
\end{align}
\Eq{eq:vca2} is an equation for the physical self-energy $\Sigma$ given the Luttinger-Ward functional $F[\Sigma]$ and the free Green's function $\GF_0$. The trace Tr is short for $\text{Tr} \equiv \frac{1}{\beta}\sum\limits_{\omega_n}\text{tr}$, the sum is over fermionic Matsubara frequencies and the small form trace tr denotes a sum over lattice sites and spin. The idea is that, due to its universality, $F[\Sigma]$ (and thus $\Omega[\Sigma, \GF_0]$) can be  evaluated exactly by exploiting a different system (so called ``reference system'') which differs from the physical system by single-particle terms only. This reference system $\hat{\mathcal{H}}'$ is simpler, and thus exactly solvable. It is defined on a cluster tiling of the original lattice and has the same interaction as the original system $\hat{\mathcal{H}}$. The VCA reference system is chosen to be a cluster decomposition of the original lattice, as the one introduced for CPT above. Comparing \eq{eq:vca3} for the full and the reference system yields 
\begin{align}
\nonumber\Omega[\Sigma, \GF_0] &= \Omega'[\Sigma, \gf_0] + \text{Tr}\;\text{ln}\left( -\left( \gf_0^{-1} -\Sigma \right) \right) \\
&- \text{Tr}\;\text{ln}\left( -\left( \GF_0^{-1} -\Sigma \right) \right)\;\mbox{,}
\label{eq:VCA2}
\end{align}
where lowercase $\gf$ denote Green's functions of the reference system. Thus the SFA/VCA approximation consists in solving \eq{eq:vca2} in a restricted range of self-energies $\Sigma$, i.e. the ones produced by the reference system. In this way, the space of allowed $\Sigma$ is spanned by the set of single-particle parameters of the reference system, $\matx'$. This means that the functional $\Omega[\Sigma, \GF_0]$ (\eq{eq:VCA2}) becomes a function of those parameters 
\begin{align}
\Omega(\matx') &= \Omega'(\matx') + \text{Tr}\;\text{ln}\left( -\GF(\matx') \right) - \text{Tr}\;\text{ln}\left( -\gf(\matx') \right)\;\mbox{,}
\label{eq:VCA3}
\end{align}
The stationarity condition determining the physical parameters \eq{eq:vca2} is then given by
\begin{align}
 \nabla_{\matx'}\Omega (\matx') &\stackrel{!}{=} 0 \; \mbox{.} 
 \label{eq:stationarity}
\end{align}
The Green's function of the physical system is obtained by the CPT relation \eq{eq:Dyson}. The matrix $\TF=\gf_0^{-1}-\GF_0^{-1}$ (\eq{eq:Dyson}) in VCA contains all single-particle terms not included in the reference system (i.e. $\TF_{\text{inter}}$) as well as, in addition, the deviation introduced by VCA, $\Delta \matx \equiv \matx'-\matx$ of the single-particle parameters of the reference system $\matx'$ with respect to the ones of the original system $\matx$. In the following we fully adopt the zero temperature formalism, in which according expressions for the grand potential and related quantities may be readily evaluated.\\ 

\subsubsection{\label{sec:SC}Alternative: Self consistent VCA} 
In \tcite{knap_nonequilibrium_2011} we explored an alternative version of VCA whereby the variational parameters $\matx'$ were determined by a suitable self consistent criterion, instead of looking for the stationary point of the grand potential $\Omega$ \eq{eq:stationarity}. This alternative approach was introduced to treat systems out of equilibrium, although it can equally be adopted in equilibrium. The advantage of this approach is that the solution of a self consistent equation is numerically easier than the search for a saddle point. The idea of this self consistent approach is to use a reference system which resembles the full system best.\\ 
The strategy is to find those values $\matx'$ for the set of parameters of the reference system which let the expectation values of their corresponding single-particle operators $\langle\hat{O}\rangle_{\text{cluster},\matx'}$ coincide with those of the full system $\langle\hat{O}\rangle_{\text{CPT},\matx,\matx'}$. Here, the angle brackets denote expectation values in the cluster and the full system coupled by CPT or VCA respectively. Consider the on-site energies $\epsilon_f'$ and $\epsilon_s'$ as variational parameters. We will look for those cluster parameters $\epsilon_f'$ and $\epsilon_s'$ which fulfill the relations
\begin{align}
\left<\hat{n}^f_{\sigma}\right>_{\text{cluster},\epsilon_f',\epsilon_s'} &\stackrel{!}{=} \left<\hat{n}^f_{\sigma}\right>_{\text{CPT},\epsilon_f,\epsilon_s,\epsilon_f',\epsilon_s'} \label{eq:SCcond}\\
\nonumber \sum_i^{L-1} \left<\hat{n}^i_{\sigma}\right>_{\text{cluster},\epsilon_f',\epsilon_s'} &\stackrel{!}{=} \sum_i^{L-1}\left<\hat{n}^i_{\sigma}\right>_{\text{CPT},\epsilon_f,\epsilon_s,\epsilon_f',\epsilon_s'}\;\mbox{.}
\end{align}
The sum is over all non interacting sites included in the cluster. This amounts to solving a system of non-linear equations in each step of the self consistency cycle. In general it is possible to vary each single-particle parameter individually. For reasons of keeping the numerics tractable we use one $\epsilon_s'$ only, which we take to be the same for each orbital in the chain. Extension to a larger number of $\epsilon_s'$ is straightforward. To fix this parameter we require the average particle density on the non interacting sites to fulfill the condition \eq{eq:SCcond}. This corresponds to the condition presented in \tcite{knap_nonequilibrium_2011} (see eq. (13) therein). In some situations, (see below,) we will alternatively consider the hybridization matrix element $V'$ and the intra-cluster hopping $t'$ as variational parameters, and proceed in an analogous way. Specifically, the particle number expectation values in \eq{eq:SCcond} are replaced by hopping expectation values. Again for $t'$ we use a single variational parameter for hopping between all uncorrelated sites and fix it by requiring the mean value of hopping in the cluster and the full system to coincide. A discussion of this self consistency condition in connection with (cluster) DMFT is given in \tcite{knap_nonequilibrium_2011}. We use an improved multidimensional Newton-Raphson algorithm to find the roots of the system \eq{eq:SCcond}. In some parameter regions no solution is found.\\
A comparison between results obtained via the usual SFA, i.e. as stationary points of the grand potential $\Omega$, which we will now refer to as \vcaomNOBLANK, to the ones obtained by the above-mentioned self consistent condition (\vcascNOBLANK) will be given in the results section (\se~\ref{sec:Results}).\\ 

\subsubsection{\label{sec:Varpar}Choice of variational parameters}
In VCA one can, in principle, optimize all possible single-particle parameters which are present in the original model, as well as additional ones. By adding bath sites not present in the original model, one includes dynamical contributions to the cluster Green's function~\cite{potthoff_self-energy-functional_2003}. The numerical difficulty increases with the number of variational parameters. For the  \vcasc case a multidimensional root finding algorithm has to be adopted. For the \vcaom case, a saddle point in many dimensions has to be located. Since the allowed set of variational parameters limits the search space for the self-energies one will find a solution in this restricted space only. It is therefore desirable to vary as many single-particle quantities as possible. A balance has to be found between a large space of available self energies and numerically feasible multidimensional algorithms. Many works have addressed the question of which parameters are the most important to vary and how the choice of variational parameters will influence or limit the results~\cite{senechal_introduction_2008}. As discussed in \tcites{aichhorn_antiferromagnetic_2006,balzer_mott_2008}, it is important to include an overall chemical potential as a variational parameter in order to preserve thermodynamic consistency. As a compromise, we will take two variational parameters $\matx=\{\epsilon_f,\epsilon_s\}$, which cover the overall chemical potential. Note that this amounts to shifting an overall on-site energy in the whole cluster plus an extra independent shift at the correlated site. For the variation of on-site energies we observe the grand potential $\Omega$ to be maximal at the stationary point which is in agreement with results for other models. Further parameters in the SIAM are the hopping $t$ and the hybridization $V$. As discussed for example in \tcite{knap_benchmarking_2010}, the variation of hopping parameters is not straightforward. For the VCA$_\Omega$ approach, we observe a maximum of $\Omega$ at $\Delta V=-V$ in the center of two symmetric stationary points. The two symmetrically lying minima are equivalent due to the fact that the self-energy is an even power of $V$. As one tunes the parameters away from particle-hole symmetry this stationary point is lost in the crossover region from the Kondo plateau to a doubly or unoccupied impurity (see \se~\ref{sec:Efscan}). In this parameter region the hopping $t$ and hybridization $V$ are probably not appropriate to be used as variational parameters within VCA$_\Omega$.\\
In the following, we always choose the set $\matx=\{V\}$ or $\matx=\{V,t\}$ for calculations at particle-hole symmetry, which also includes $\{\epsilon_f,\epsilon_s\}$, since the variation of on-site energies will always yield zero deviations from the physical parameters and thus reproduce the CPT result here. For all other parameter regions it is sufficient to consider $\matx=\{\epsilon_f,\epsilon_s\}$ as variational parameters.\\

\section{\label{sec:Omega}Grand Potential for reference systems of infinite size}
The reference system consists of two parts, a finite interacting system and a non interacting system of infinite size, the environment.  $\Omega'(\matx')$ is given by the sum of the grand potentials of the interacting cluster ($\Omega'_{\text{interacting}}$) and of the environment ($\Omega'_{\text{env}}$) (, which correspond to $\hat{\mathcal{H}}_{\text{interacting}}$ (\eq{eq:cluster}) and $\hat{\mathcal{H}}_{\text{environment}}$ (\eq{eq:environment})). Here we outline how to determine the grand potential for such kinds of reference systems. For the Green's function \GFBLANK within the CPT/VCA approximation the Dyson equation is given in \eq{eq:Dyson}. The Green's function and \TFBLANK have the block structure
\begin{align*}
\GF &=
\begin{pmatrix}
\GF_{cc} & \GF_{ce}\\
\GF_{ec}&\GF_{ee}
\end{pmatrix} \,\mbox{, }
&\TF &=
\begin{pmatrix}
\TF_{cc} & \TF_{ce}\\
\TF_{ec}&0
\end{pmatrix} \,\mbox{.}
\end{align*}
Up to this point all matrices involving environment indices have infinite size. As far as the Green's function itself is concerned this is no problem as we are primarily interested in $\text{G}_{cc}$ for which the Dyson equation reduces to
\begin{align*}
\GF_{cc} &= 
\gf_{cc} + \gf_{cc} \TF_{cc} \GF_{cc}+ \gf_{cc} \TF_{ce} \GF_{ec}\\
\GF_{ec} &= \gf_{ee} \TF_{ec} \GF_{cc}\;\mbox{,}
\end{align*}
and therefore
\begin{align*}
\GF_{cc} &= 
\gf_{cc} + \gf_{cc} \tilde \Sigma_{cc} \GF_{cc}\\
\tilde\Sigma_{cc} &:= \TF_{cc} + \TF_{ce} \gf_{ee} \TF_{ec}\;\mbox{.}
\end{align*}
A bit more tedious is the elimination of the explicit dependence on the environment part of G, as far as the grand potential \eq{eq:VCA3} is concerned. We start out from a form of the grand potential functional given by S\'en\'echal~\cite{senechal_introduction_2008}
\begin{align}\label{eq:aux0}
\Delta \Omega &:= \Omega-\Omega' =- \Tr\ln \bigg(\uu - \TF \gf\bigg)\;.
\end{align}
In \app~\ref{sec:AppendixOmeganew} it is shown that $\Delta\Omega$ can be expressed solely in terms of cluster quantities 
\begin{align}\label{eq:aux3}
\Delta\Omega &= - \Tr\ln \bigg(
\uu_{cc} - \tilde\Sigma_{cc} \gf_{cc}
\bigg)\;.
\end{align}
Along the lines outlined in \tcite{senechal_introduction_2008}, the resulting integral can be regularized and expressed as
\begin{align}
\label{eq:omegaResult}
{\Omega - \Omega'_{\text{env}}} &=\, \Omega'_{\text{interacting}} + \text{tr}\left( \TF\right) \\
&- \frac{1}{\pi} \sum_{\sigma} \int_{0}^{\infty}
d\omega\text{ln}\left|\text{det}\left(\uu_{cc} - 
\tilde\Sigma_{cc}^\sigma(i\omega) \gf_{cc}^\sigma(i\omega)\right)\right|\,\mbox{.}\nonumber
\end{align}
The quantities $\Omega'$ are the grand potentials of the uncoupled reference system. The constant infinite contribution $\Omega'_{\text{env}}$ is absorbed into the definition of $\Omega$. It plays no further role as it does not depend on the variational parameters. This integral may be evaluated as suggested in \tcite{senechal_introduction_2008} by integrating from 0 to $\Lambda_1$, from $\Lambda_1$ to $\Lambda_2$ and from $\Lambda_2$ to $\infty$. $\Lambda_1$ and $\Lambda_2$ represent two characteristic scales in the problem (for example the smallest/largest eigenvalue of the Hamiltonian matrix). For the last part of the integral a substitution $\tilde{\omega}=\frac{1}{\omega}$ is performed. We use an adaptive Gauss Legendre integrator for the evaluation.\\

\section{\label{sec:Results}Results}
We have evaluated several benchmarking dynamic quantities of the SIAM. In the following, results for the impurity density of states will be presented and compared to ED, NRG and DMRG data. We will elaborate on the strengths and weaknesses of the methods as well as the comparison of CPT to VCA. Furthermore, we will discuss the relation between VCA$_{\text{SC}}$, where the variational parameters are determined self consistently via \eq{eq:SCcond} and VCA$_{\Omega}$, where the variational parameters are defined at the stationary point of the grand potential. We will show that the Kondo resonance is reproduced within the framework of CPT/VCA and that the variational results fulfill certain analytic relations like the Friedel sum rule (\eq{eq:FSR}). The method will be shown to provide reasonably accurate results in a wide range of parameter regimes of the model. Low energy properties related to the Kondo temperature $T_K$ will be discussed in context with renormalization group results. The imaginary frequency Green's function and self-energy will be compared to CT-QMC results.\\

\subsection{\label{sec:OddOdd}Even-Odd Effect - choice of the impurity position}
CPT/VCA rely on the Green's function of an interacting cluster of size $L$ which is obtained by exact diagonalization. Due to this fact, it is unavoidable that some effects of the finite size cluster influence the solution of the full system. (Except in the case of vanishing interaction strength, i.e. $U=0$.) Therefore suitable clusters have to be chosen on a basis of physical results. Some aspects of this are discussed by Balzer \etal~\cite{balzer_mott_2008} in the context of DMFT and VCA and by Hand \etal~\cite{hand_spin_2006} in the context of DMRG. In this work we consider interacting clusters of even size only. For these systems the ground state does in general not suffer from spin degeneracy. Furthermore, the spatial position of the impurity is important. This can be inferred from the bath's density of states, which vanishes for $\omega=0$ at every second site. It may also be seen in the structure of the ground state, for which the size of the degenerate sectors alternates with the geometrical size of the cluster. Throughout this work we position the impurity f-orbital at the beginning of the infinite chain, although essentially the same results are achieved by attaching it to an s-orbital at site two, four, etc. inside the chain.\\

\subsection{\label{sec:SpectralProperties}Spectral properties}
\begin{figure}
        \centering
        \includegraphics[width=0.48\textwidth]{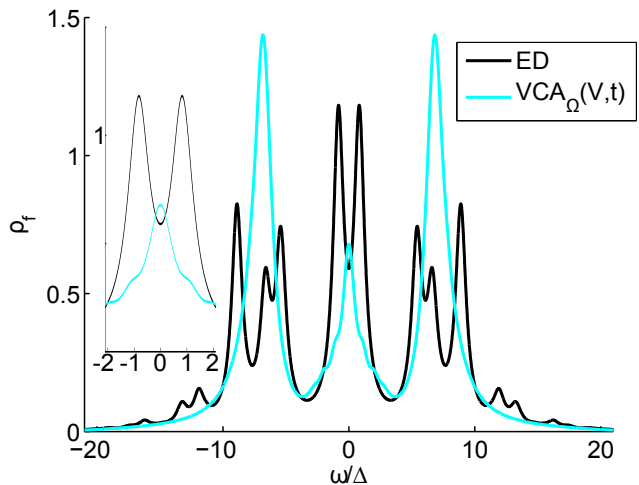}
        \caption{(Color online) Comparison of the local single-particle spectral function at the impurity f-orbital at particle-hole symmetry as obtained by exact diagonalization (ED) of a ten site chain (black) and \vcaom (cyan). \vcaom was used with two variational parameters: the hopping $t$ and the hybridization $V$ considering a length of the interacting part of the reference system of $L=10$. All data shown is for an interaction strength of $U/\Delta=12$. All results have been obtained for a large numerical broadening $0^+=0.05$. The inset shows a zoom to the low energy region.}
        \label{fig:ED_CPT_VCA}
\end{figure}

\begin{figure*}
        \centering \includegraphics[width=0.98\textwidth]{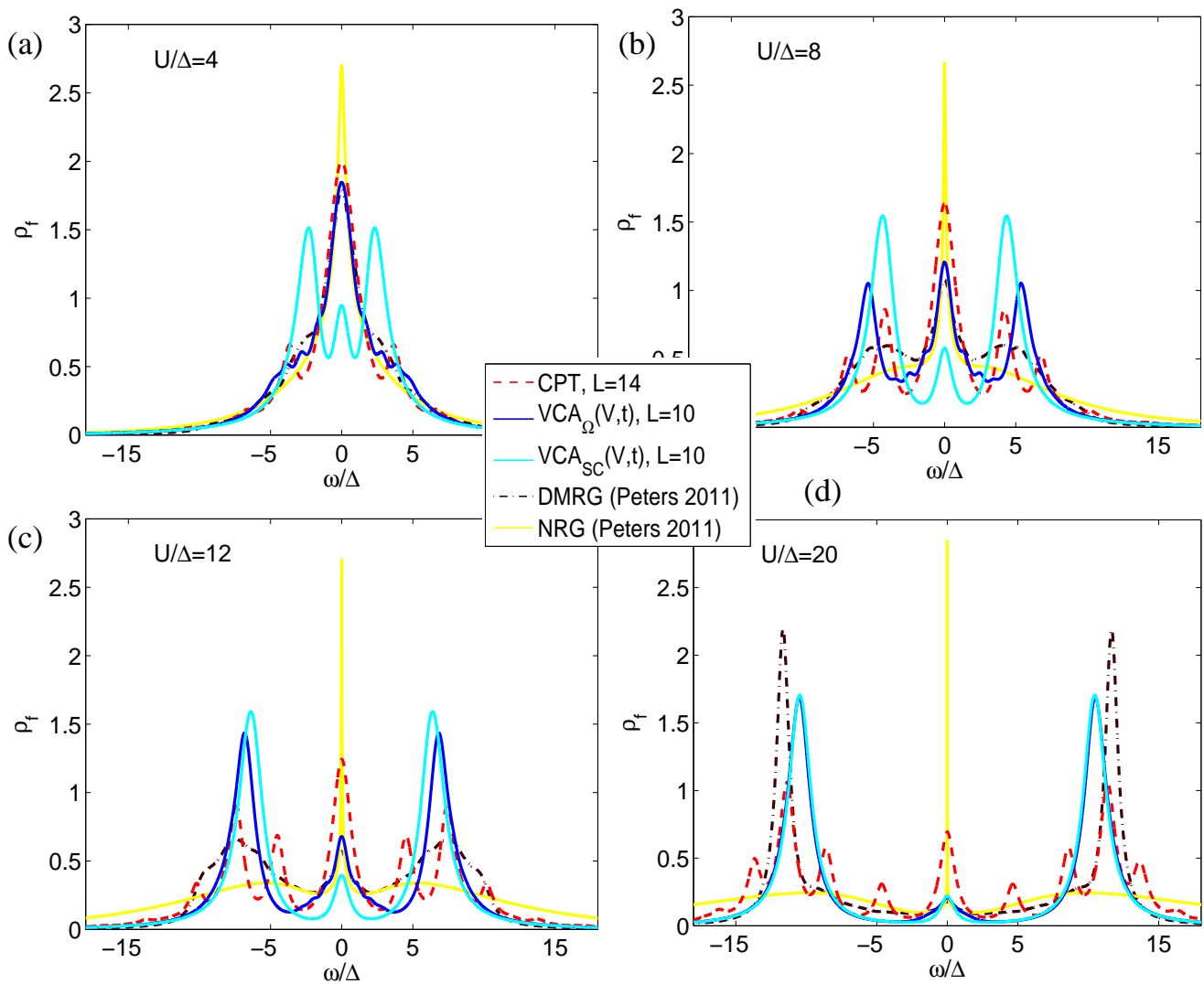}
        \caption{(Color online) Single-particle spectral function at the impurity f-orbital at particle-hole symmetry for different interaction strengths $U$. The interaction strengths shown are $U/\Delta=4$ in the upper left figure (a), $U/\Delta=8$ in the upper right figure (b), $U/\Delta=12$ in the lower right figure (c) and $U/\Delta=20$ in the lower right figure (d). Each plot shows the results obtained by CPT for a length of the interacting part of the reference system of $L=14$ (dashed-red), VCA$_\Omega$ with two variational parameters: the hopping $t$ and the hybridization $V$ which are determined by the stationary point of the grand potential $\Omega$ at a length of the interacting part of the reference system of $L=10$ (blue), VCA$_{\text{SC}}$ with the same variational parameters determined self consistently at a length of the interacting part of the reference system of $L=10$ (cyan). All results have been obtained for a large numerical broadening $0^+=0.05$. As a reference the NRG and DMRG results of Peters~\cite{peters_spectral_2011} are plotted in yellow and dash-dotted-dark brown respectively.} 
        \label{fig:spectraPHsym}
\end{figure*}

\begin{figure}
        \centering
        \includegraphics[width=0.48\textwidth]{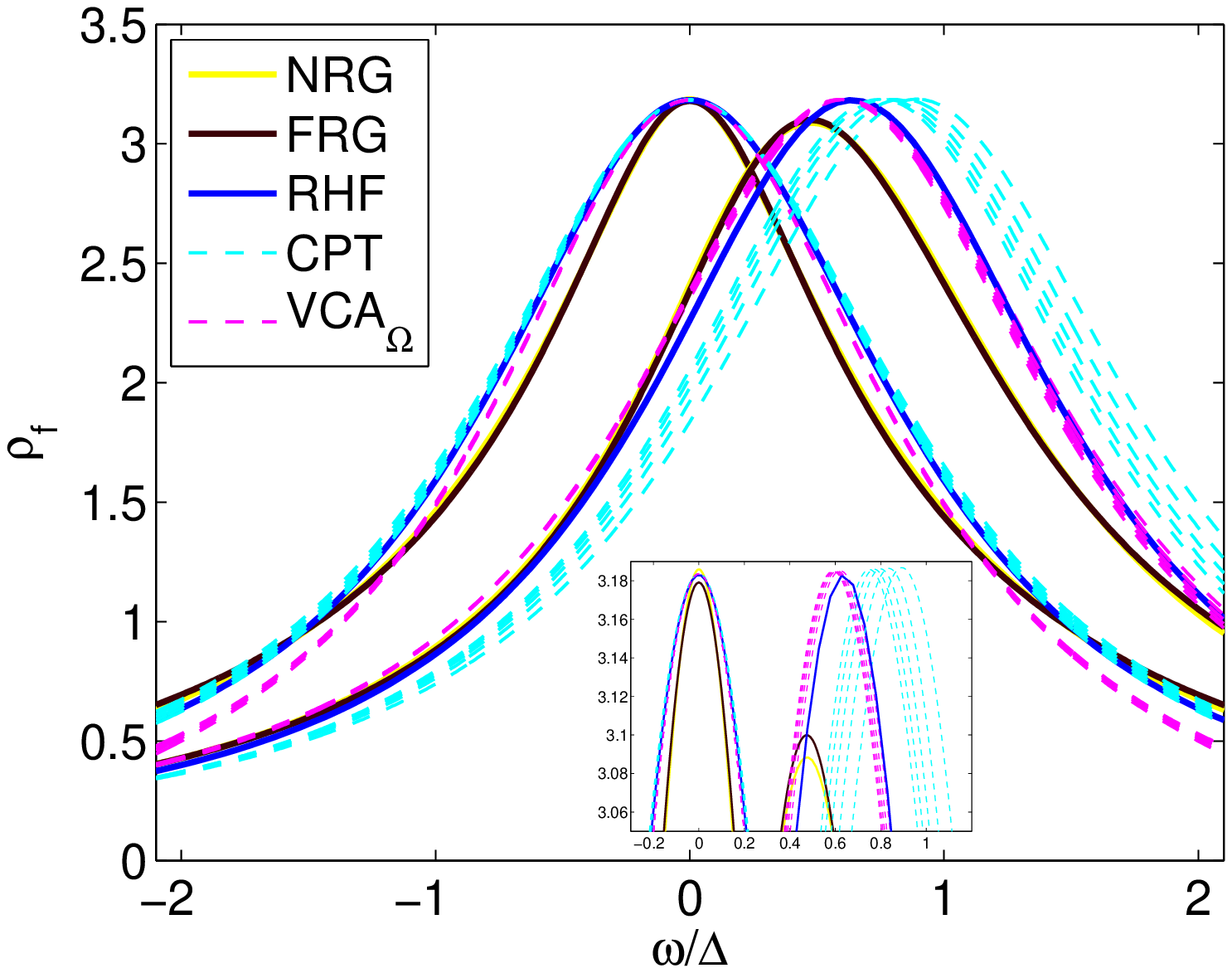}
        \caption{(Color online) Magnification of the Kondo resonance in the density of states of the impurity f-orbital. Shown are calculations for two different sets of parameters. The resonance at $\omega = 0$ corresponds to the particle-hole symmetric model: $U/\Delta=20$, $\epsilon_f/\Delta=-10$, while the resonance away from zero corresponds to a set of parameters right at the edge of the Kondo region: $U/\Delta=20$, $\epsilon_f/\Delta=0$. For comparison we show NRG (yellow) and FRG (dark brown) data as well as results obtained from a restricted Hartree-Fock calculation (blue) from Karrasch \etal~\cite{karrasch_finite-frequency_2008} (The NRG results are partially hidden by the FRG results.). The CPT result (cyan) is shown for lengths of the interacting part of the reference system $L={2, 4, 6, 8 \mbox{ and } 10}$. Results for higher $L$ are always located towards the center of the plot. In the particle-hole symmetric case \vcaom (magenta) was performed with variational parameters $\matx=\{V\}$ for $L={2,4,6,8 \mbox{ and } 10}$. Away from particle-hole symmetry \vcaom was performed with variational parameters $\matx=\{\epsilon_f,\epsilon_s\}$ for the same lengths of the interacting part of the reference system $L$. For the CPT/VCA calculations a numerical broadening of $0^+=10^{-6}$ was used. The inset shows a zoom to the top region of the peaks.}
        \label{fig:spektralInset}
\end{figure}

\begin{figure}
        \centering
        \includegraphics[width=0.48\textwidth]{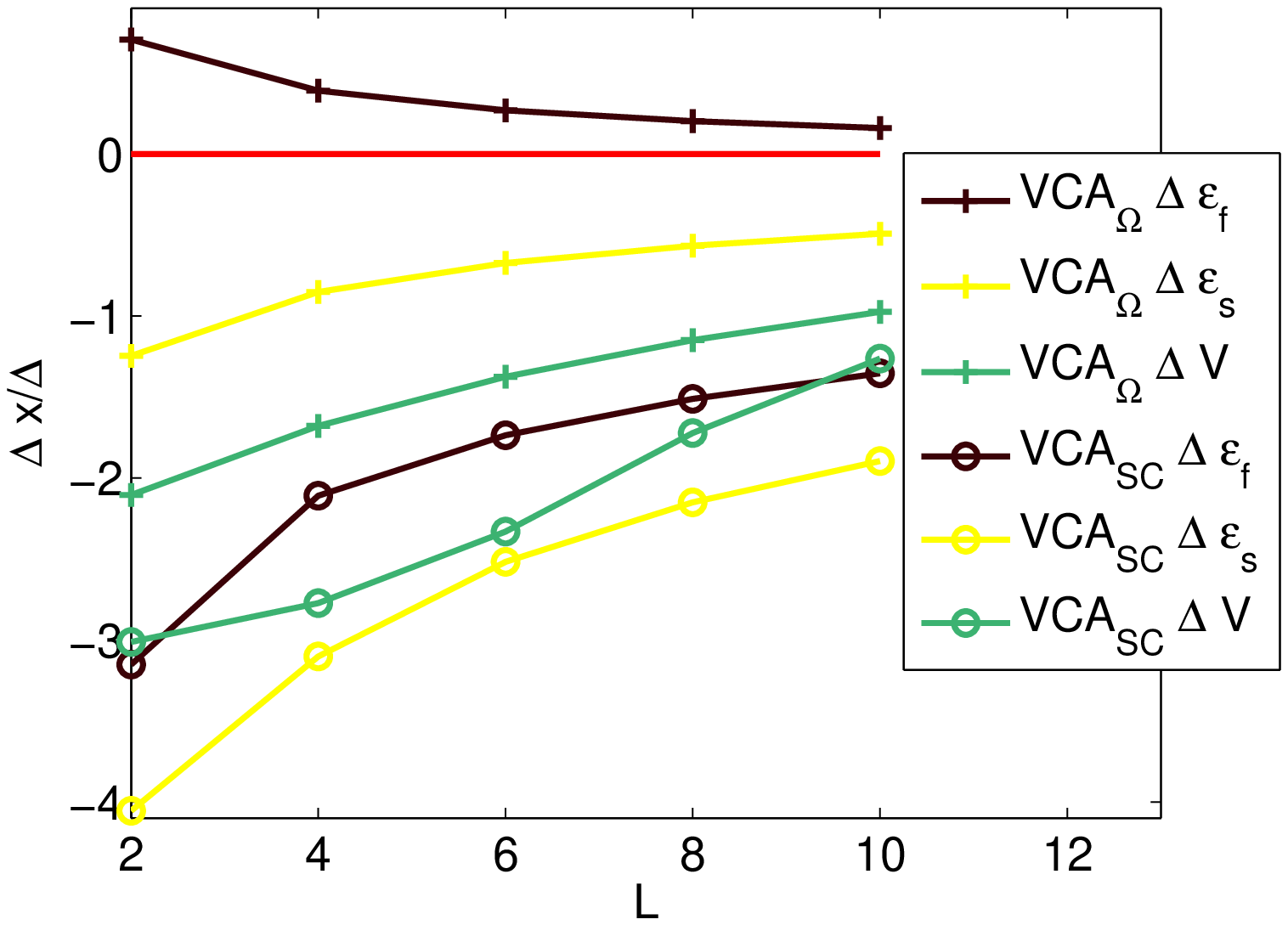}
        \caption{(Color online) Evolution of the variational parameters for the data shown in \fig{fig:spektralInset}. Shown is the difference of the parameters of the reference system $\matx'$ to the physical parameter $\matx$: $\Delta \matx = \matx' - \matx$. Parameters obtained by \vcaom (crosses) are compared to those obtained by \vcasc (circles). The variational parameters $\Delta \epsilon_f$ (dark brown) and $\Delta \epsilon_s$ (yellow) correspond to the calculation away from particle-hole symmetry in \fig{fig:spektralInset} while the parameter $\Delta V$ (olive) corresponds to the calculation at particle-hole symmetry. Lines are only guides to the eye.} 
        \label{fig:varPar}
\end{figure}

The single-particle spectral function $A_{ii}^{\sigma}$ is obtained from the retarded Green's function $\,\GF_{ii}^{\sigma,\text{ret}}$ (see e.g. \tcite{negele_quantum_1998})
\begin{align}
  A_{ii}^\sigma(\omega) &= -\frac{1}{\pi}\,\Im{\text{m}}\left(\GF_{ii}^{\sigma,\text{ret}}(\omega)\right) \;\mbox{.}
\label{eq:spectral}
\end{align}
The diagonal element at the impurity f-orbital $A_{ff}^\sigma(\omega)$ describes the impurity density  of states $\rho_{f}^\sigma(\omega)$. A physical property of the SIAM which poses a challenge to numerical methods is the Kondo-Abrikosov-Suhl resonance often referred to as Kondo peak~\cite{kondo_resistance_1964}. It arises in the parameter regime where the magnetic moment of the impurity is screened by the conduction electrons to form a singlet state~\cite{coleman_local_2002}. The particle-hole symmetric model lies in the center of this Kondo region. This quasi particle excitation is for example not captured in mean field approaches. With increasing interaction strength $U$ the numerical solution becomes increasingly challenging.\\
In this section we elaborate on the results for the density of states in the particle-hole symmetric case.\\
A comparison of the local single-particle spectral function at the impurity f-orbital as obtained by exact diagonalization and \vcaom is shown in \fig{fig:ED_CPT_VCA}. The ED result is for a ten site system with open boundary conditions. The \vcaom result is for an infinite reference system, where the interacting part of the reference system was taken to be of size $L=10$. An ED treatment of a finite-size SIAM can not reproduce the low energy resonance at zero energy in the single-particle spectral function (see also \app~\ref{sec:AppendixFSR}). It consists, in the particle-hole symmetric case, for an even number of sites (and open boundary conditions), of symmetrically lying excitations which shift closer to zero for increasing system size and represent a large energy scale. For an odd number of sites a pole in the local single-particle Green's function of the impurity f-orbtial will be present at $\omega=0$. CPT as well as VCA are able to reproduce finite spectral weight at $\omega=0$ (even for $0^+\rightarrow 0$), since these methods are formulated for an infinite system. The finite-size structure in the high energy incoherent part of the spectrum, owing from the excitations of the interacting part of the reference system, is strongly reduced in \vcaomNOBLANK~.\\
Results for the single-particle spectral function \eq{eq:spectral} of the impurity f-orbital are shown in \fig{fig:spectraPHsym} for four different interaction strengths $U/\Delta=4,8,12 \mbox{ and } 20$. As a reference, the spectra obtained with NRG and DMRG from Peters~\cite{peters_spectral_2011} are plotted. Renormalization group approaches like NRG are especially suited to reproduce the low energy quasi particle excitations of this model and therefore serve as a reference for our data. The spectra of Peters were obtained for a flat conduction electron density of states, which was mapped by linear discretization in energy space onto the corresponding orbitals of a semi-infinite chain. Our model is based on a semi-circular density of states of the conduction electrons. The low energy part of the spectra is comparable because we have chosen the only relevant parameter for the low energy part of the spectrum: $\Delta$ accordingly. This parameter fully determines the influence of the conduction electrons on the impurity f-orbital for low energies and therefore the low energy part of the spectrum. The high energy part of the spectrum may deviate slightly and is not directly comparable but yields a crude reference. In addition, we have chosen here a very large numerical broadening of $0^+=0.05$ for reasons of comparison only. This value was used in the DMRG calculations and is needed there to obtain spectra using the correction vector method. This influences the width and the height of the Kondo resonance, located at $\omega=0$. The CPT spectral weight at $\omega=0$ appears too broad in the plot in comparison with the NRG result. This is only partly due to a large numerical broadening. Due to the nature of the CPT method we cannot expect it to reproduce the low energy spectrum as well as RG calculations do. The height of the Kondo resonance appears too small in this figure because of the large $0^+$. It converges with $0^+\rightarrow 10^{-6}$ to the result predicted by scattering theory (see \fig{fig:spektralInset} and \se~\ref{sec:Efscan}). The high energy incoherent parts of the spectrum located at $\omega \approx -\epsilon_f$ and $\omega \approx -\epsilon_f+U$ develop more and more with increasing length of the interacting part of the reference system $L$. A comparison of the center of gravity of the high energy incoherent part of the spectrum of the $L=14$ site CPT result and the fifty site DMRG result is in reasonable agreement. There are spurious structures in the spectral density, originating from the cluster Green's function of the finite system, preventing continuous spectra to form. We would like to note that the accurate determination of the Green's function of the reference system is of prime importance. An inaccuracy in pole-positions or pole-weights for very small but non-vanishing weights will yield spurious artifacts in the spectra in the vicinity of $\omega=0$.\\
To improve on the result of CPT we considered the hopping matrix element $t$ and the hybridization matrix element $V$ as variational parameters. The parameters used for the evaluation of the reference system were determined with two different methods. \vcaom results are depicted in the plot for a length of the interacting cluster of $L=10$. As shown in the figure this method strongly reduces the finite size peaks in the high energy incoherent part of the spectrum. The width of that part of the spectrum is reproduced correctly for high values of $U$ where the full width at half maximum (FWHM) within VCA is given by $\approx 1.9\Delta$. This comes very close to the expected $2\Delta$~\cite{brenig_theory_1974,gunnarsson_handbook_1987} of the high energy atomic excitations. VCA improves the spectral properties of the Kondo resonance with respect to CPT, bringing it closer to the fifty site DMRG result. The data obtained using the self consistent VCA approach \vcasc agree very well with the result based on \vcaomNOBLANK on the position of the spectral features. The respective weight however disagrees for low values of interaction strength $U$, which is due to the different values predicted for the variational parameters by the two procedures. One should note that the two broad Lorentzian high energy peaks (in \vcaom as well as \vcascNOBLANK) consist of many excitations which will be revealed upon repeating this calculation with smaller $0^+$.\\
\begin{figure}
        \centering
        \includegraphics[width=0.48\textwidth]{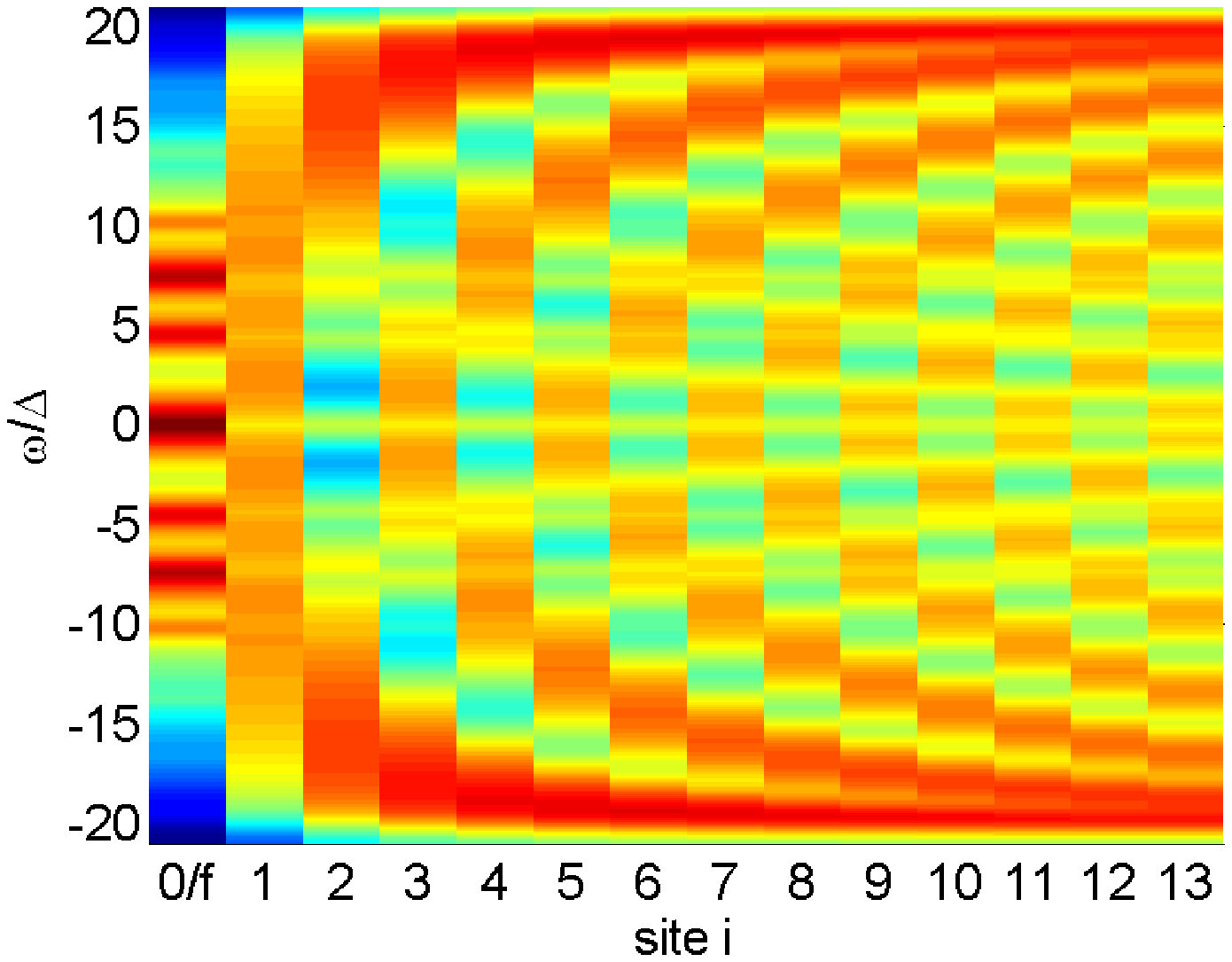}
        \caption{(Color online) The local density of states \eq{eq:spectral} is shown resolved in real space. The spectrum was obtained using CPT on a $L=14$ site interacting cluster. The impurity parameters were $U/\Delta=12$, $\epsilon_f/\Delta=-6$ and the numerical broadening was set to $0^+=0.05$. The spectrum shown in \fig{fig:spectraPHsym} (c) corresponds to the data shown for the impurity f-orbital located at site $0$ in the plot. The density plot is shown with a logarithmically scaled coloring from blue indicating zero to red indicating high values.}
        \label{fig:SpatialSpectrum}
\end{figure}
A more detailed look on the spectral region of the Kondo resonance is provided in \fig{fig:spektralInset}. The CPT/VCA$_\Omega$ data is compared to NRG and FRG data as well as results obtained from a restricted Hartree-Fock calculation from Karrasch \etal~\cite{karrasch_finite-frequency_2008}. The CPT/VCA results are plotted for lengths of the interacting part of the reference system $L={2,4,6,8 \mbox{ and } 10}$ for two different sets of parameters. The results for higher $L$ are always located towards the center of the figure. The results corresponding to the resonance at $\omega=0$ were obtained for the particle-hole symmetric model. For this set of parameters we used the hybridization $V$ as a variational parameter. The second peak shown centered around $\omega/\Delta\approx0.8$ corresponds to a parameter set right at the border of the Kondo region. The variational parameters used away from particle-hole symmetry are $\matx=\{\epsilon_f,\epsilon_s\}$. One can see that the CPT result is not converged for $L=10$ site interacting clusters yet. In contrast, the VCA$_\Omega$ result seems to converge much faster. Although in the plot it looks like the VCA result does not improve much upon a restricted Hartree Fock calculation, we will show in the following that CPT/VCA yields results in all parameter regimes of the SIAM which cannot be reproduced within a mean field treatment.\\ 
The variational parameters obtained for the two sets of parameters shown in \fig{fig:spektralInset} are presented in \fig{fig:varPar}. In addition to the VCA$_\Omega$ parameters, which were used for the results above, the variational parameters obtained in \vcasc are also depicted. We plotted the difference of the parameter of the reference system $\matx'$ to the physical parameter $\matx$: $\Delta \matx$. All deviations $\Delta \matx$ appear to converge to zero with increasing length $L$ of the interacting part of the reference system. Notice that the self consistent approach always leads to a $\Delta \matx$ of greater magnitude with respect to \vcaomNOBLANK. Remarkably, the spectra obtained by \vcaom and \vcasc for the parameter set $\matx=\{\epsilon_f,\epsilon_s\}$ are in very good agreement even though the variational parameters are rather different. The most striking difference is that the self consistent approach yields a negative $\Delta \epsilon_f$ while the $\Omega$ based VCA yields a positive $\Delta \epsilon_f$. This is however compensated by the different $\Delta \epsilon_s$. Using the hybridization $V$ as a variational parameter, the $\Delta V$ obtained by VCA$_\Omega$ and VCA$_{\text{SC}}$ agree rather well. Remarkably, the resulting density of states is very different, which shows that the calculation is extremely sensitive to this parameter.\\
A spatially resolved image of the spectral function, calculated with CPT, for the parameter set used in \fig{fig:spectraPHsym} (c) is shown in \fig{fig:SpatialSpectrum}. The qualitative picture would be the same in VCA, merely the structures are slightly shifted. This view reveals how the perturbation, introduced by the impurity, is fading away slowly in an alternating fashion. At every second site away from the impurity a dip at $\omega=0$ is present, which is usually referred to as Fano dip.\\
The low computational effort of CPT/VCA proofs advantageous for calculating spectra. The VCA procedure (for a twelve site interacting cluster) usually converges in minutes to hours on a standard workstation PC, while more demanding numerical methods often need days to a week to converge. Furthermore, the spectra are exactly determined from the Lehmann representation and no ill-posed analytical continuation is required in comparison to methods working in imaginary time or imaginary frequency space. To our knowledge the most accurate spectra available for this model so far are published in \tcite{zitko_energy_2009}.\\ 
Overall one may conclude that CPT, \vcaom and \vcasc reproduce a Kondo resonance, which fulfills the Friedel sum rule (\eq{eq:FSR}) for $0^+\rightarrow10^{-6}$. The VCA results improve drastically upon the CPT data which may be seen in a much faster convergence in $L$ and a suppression of finite size effects especially in the high energy part of the spectrum which in addition has the expected width within VCA. \vcaom and \vcasc agree rather well on the position of the spectral features. However they assign very different spectral weight to them at low values of interaction strength $U$.\\

\subsection{\label{sec:Efscan}Impurity density of states and occupation}
\begin{figure}
    \centering
        \includegraphics[width=0.48\textwidth]{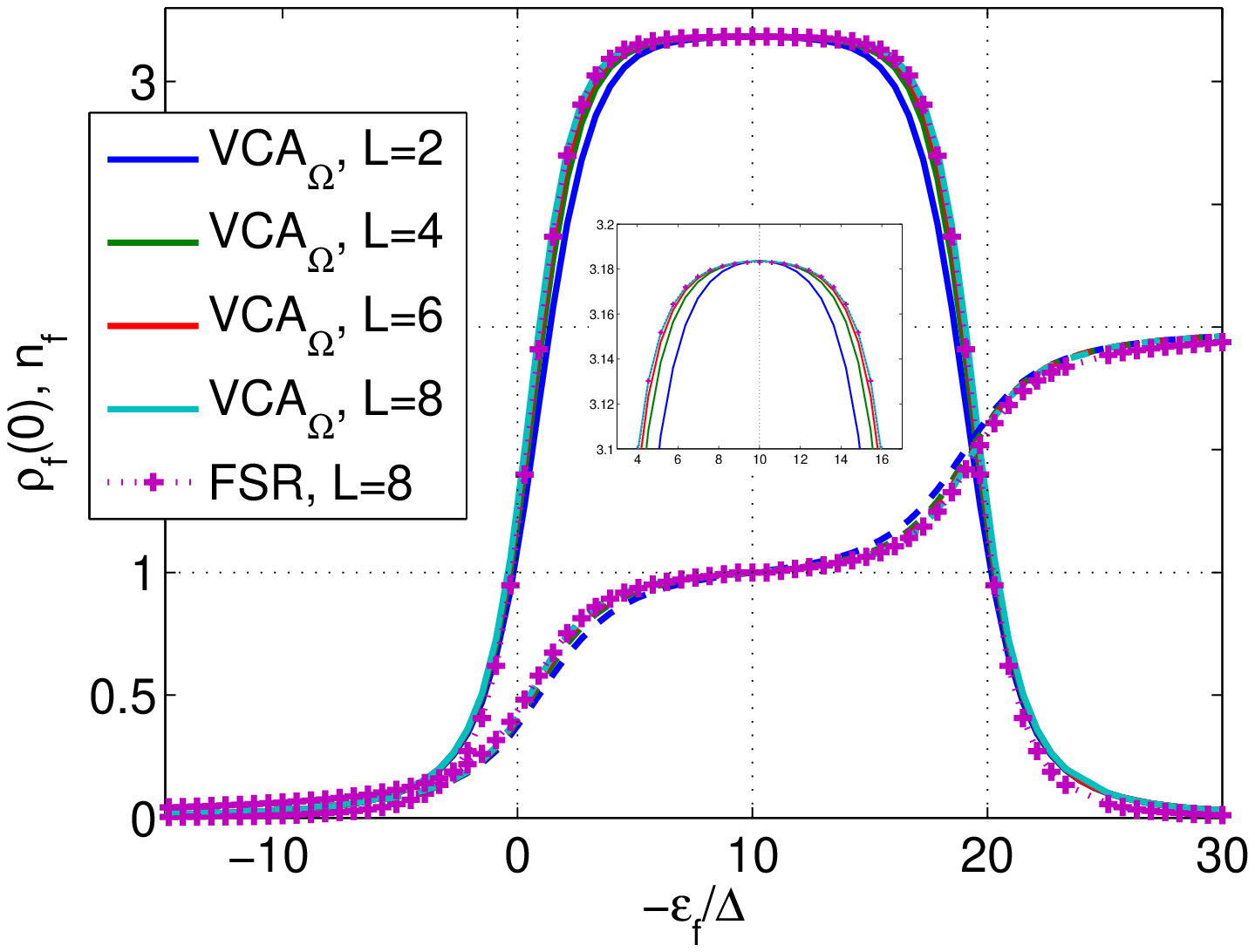}
        \caption{(Color online) Density of states of the impurity f-orbital (solid lines)
obtained via \vcaom at $\omega=0$ and average occupation of the impurity (dashed lines) for different lengths of the interacting part of the reference system $L = 2, 4, 6 \mbox{ and } 8$ (blue, green, red and cyan) as a function of the impurity on-site energy $\epsilon_f$. The Coulomb interaction $U$ is kept constant at $U/\Delta=20$. The numerical broadening used is $0^+=10^{-6}$. The set of single-particle parameters considered for variation within VCA$_{\Omega}$ is $\matx=\{\epsilon_f,\epsilon_s\}$. Note that here the point $\epsilon_f=-\frac{U}{2}$ corresponds to the particle-hole symmetric case. The Friedel sum rule (\eq{eq:FSR}) was applied to the $L=8$ result (dotted-violet). It is fulfilled to a very good approximation in the Kondo region and far outside of it. Small deviations from the Friedel sum rule arise at the crossover region to an empty or doubly occupied impurity. The inset shows a zoom to the Kondo plateau.}
        \label{fig:efScanVCA}
\end{figure}

\begin{figure}
        \centering
        \includegraphics[width=0.48\textwidth]{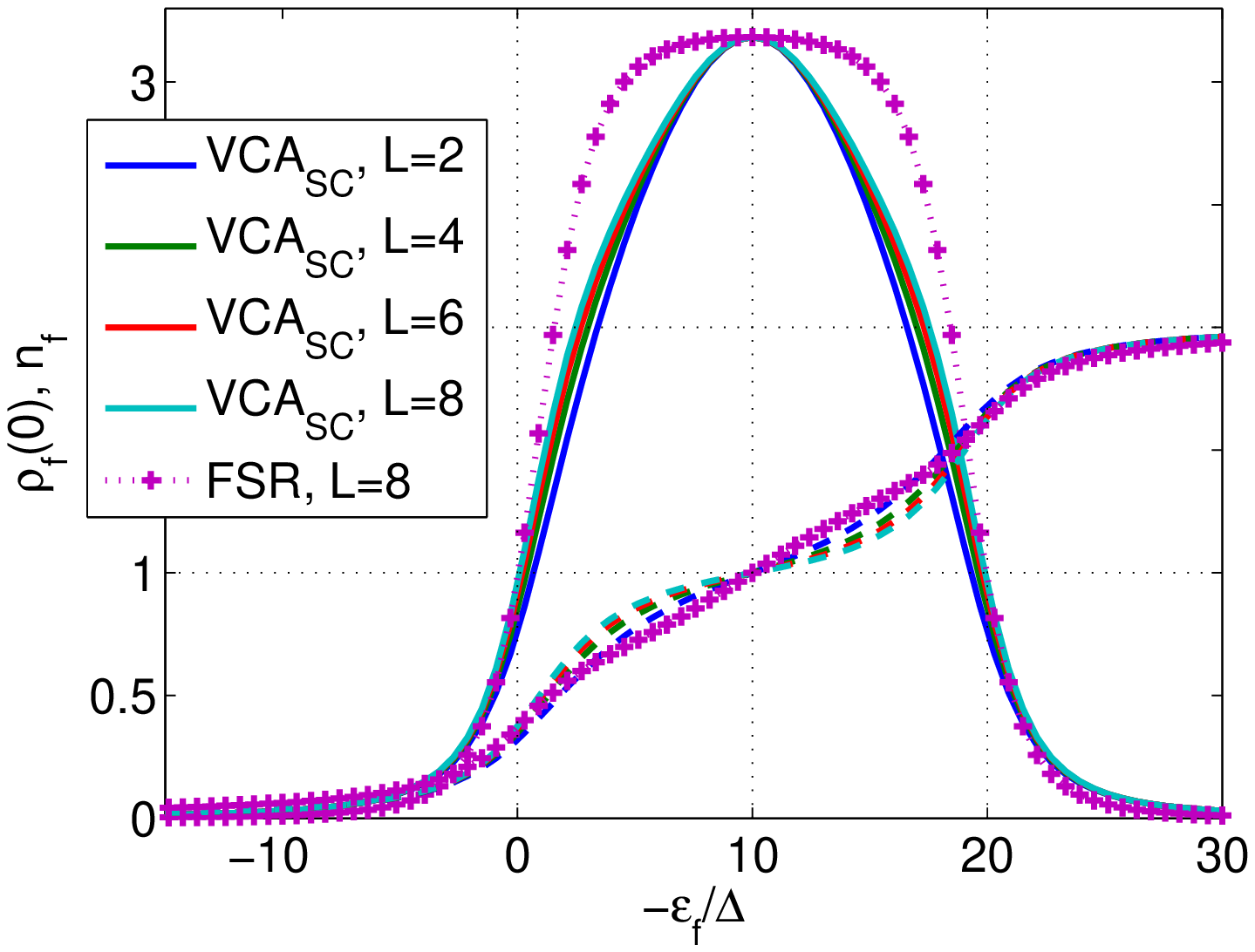}
        \caption{(Color online) Density of states of the impurity f-orbital (solid lines) obtained via \vcasc at $\omega=0$ and average occupation of the impurity (dashed lines) for different lengths of the interacting part of the reference system $L = 2, 4, 6 \mbox{ and } 8$ (blue, green, red and cyan) as a function of the impurity on-site energy $\epsilon_f$. The Coulomb interaction $U$ is kept constant at $U/\Delta=20$. The numerical broadening used is $0^+=10^{-6}$. The set of single-particle parameters considered for variation within \vcasc is $\matx=\{\epsilon_f\}$. Note that here the point $\epsilon_f=-\frac{U}{2}$ corresponds to the particle-hole symmetric case. The Friedel sum rule (\eq{eq:FSR}) was applied to the $L=8$ result (dotted-violet). It is fulfilled in a region of $n_f\lesssim0.4$ and $n_f\gtrsim1.6$.}
        \label{fig:efScanSCN}
\end{figure}

\begin{figure}
        \centering
        \includegraphics[width=0.48\textwidth]{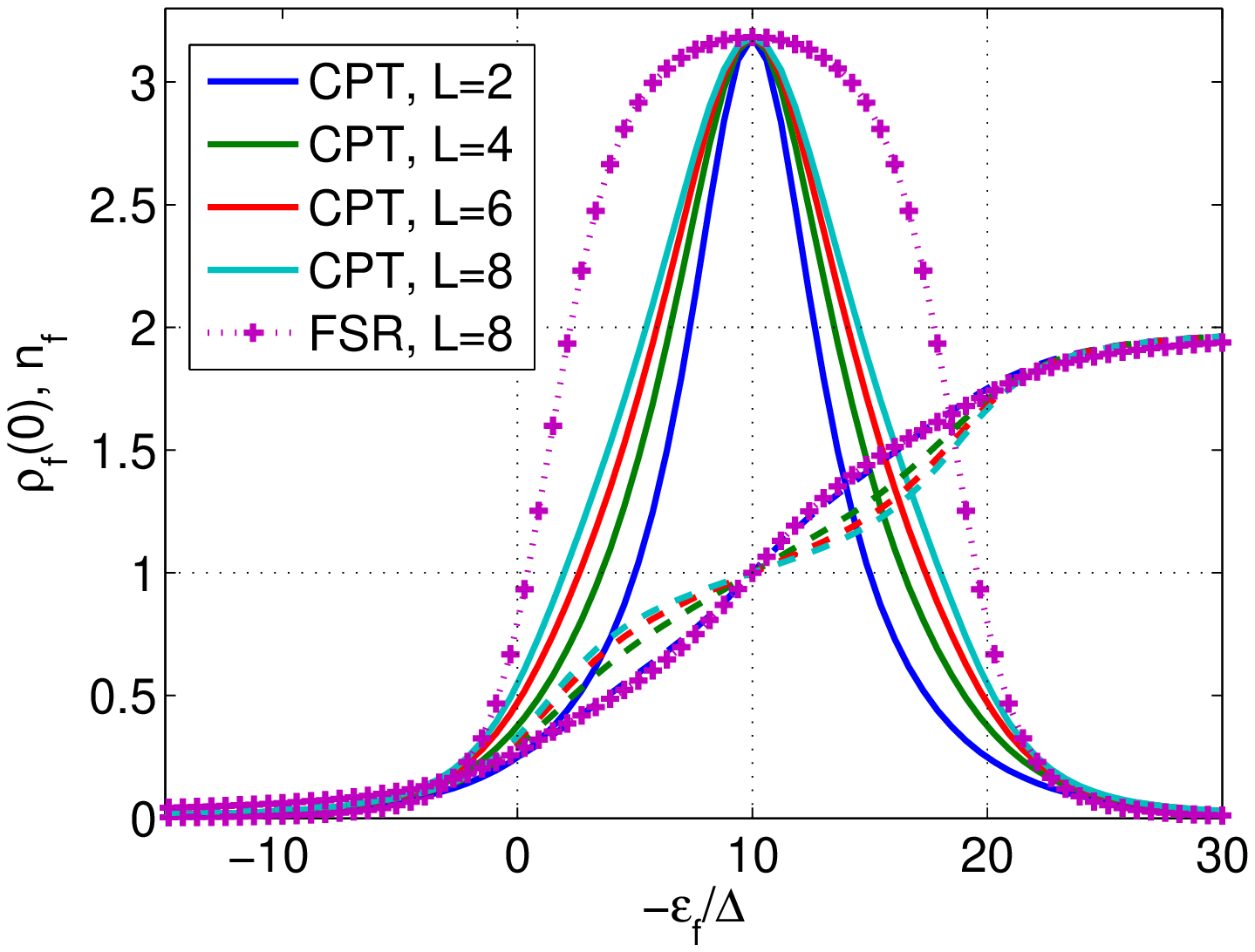}
        \caption{(Color online) Density of states of the impurity f-orbital (solid lines) obtained via CPT at $\omega=0$ and average occupation of the impurity (dashed lines) for different lengths of the interacting part of the reference system $L = 2, 4, 6 \mbox{ and } 8$ (blue, green, red and cyan) as a function of the impurity on-site energy $\epsilon_f$. The Coulomb interaction $U$ is kept constant at $U/\Delta=20$. The numerical broadening used is $0^+=10^{-6}$. Note that here the point $\epsilon_f=-\frac{U}{2}$ corresponds to the particle-hole symmetric case. The Friedel sum rule (\eq{eq:FSR}) was applied to the $L=8$ result (dotted violet). It is drastically violated. However the results are far from converged for the small lengths of the interacting part of the reference system considered here.}
        \label{fig:efScanCPT}
\end{figure}

\begin{figure}
        \centering
        \includegraphics[width=0.48\textwidth]{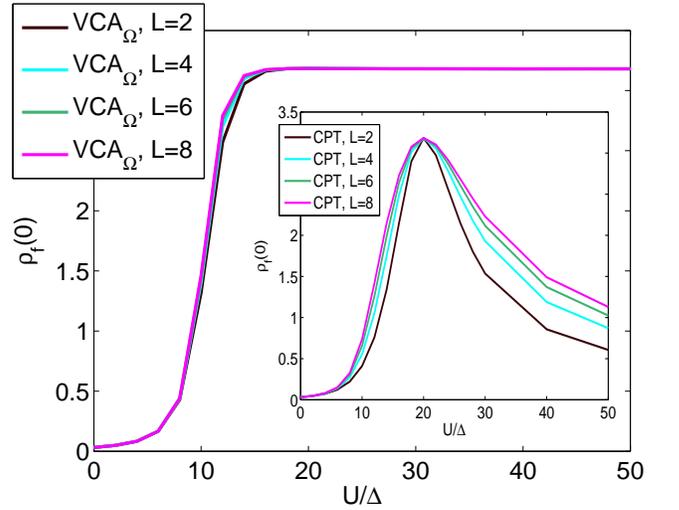}
        \caption{(Color online) Density of states of the impurity f-orbital at $\omega=0$ for different lengths of the interacting part of the reference system $L=2, 4, 6 \mbox{ and } 8$ (dark brown, cyan, olive and magenta) as a function of the interaction strength $U$. The impurity on-site energy $\epsilon_f$ is kept constant at $\epsilon_f/\Delta=-10$. The numerical broadening is chosen to be $0^+=10^{-6}$. The set of single-particle parameters considered for variation within \vcaom is $\matx=\{\epsilon_f,\epsilon_s\}$. The inset shows the CPT results.}
        \label{fig:UScan}
\end{figure}

The occupation of the impurity f-orbital is given at $T=0$ by
\begin{align}
  \langle n_{\sigma}^f\rangle &= \frac{1}{2} + \frac{1}{\pi}\int_0^\infty\,d\omega\,\Re{\text{e}}\left(\GF_{ff}^\sigma(i\omega)\right)\;\mbox{.}
\label{eq:n}
\end{align}
This integral may be evaluated from the imaginary frequency Green's function, which in turn is directly accessible within CPT/VCA.\\
To see whether CPT/VCA are good approximations in all parameter regions of the SIAM, we vary the on-site energy of the impurity $\epsilon_f$ at fixed interaction strength $U$. The local impurity density of states at the chemical potential ($\omega=\mu=0$) and the impurity occupation number are plotted for various lengths of the interacting part of the reference system $L=2, 4, 6 \mbox{ and } 8$ for the same model parameters. The \vcaom result is shown in \fig{fig:efScanVCA}, a \vcasc calculation in \fig{fig:efScanSCN} and the CPT data in \fig{fig:efScanCPT}.\\
We start out by discussing the \vcaom result (\fig{fig:efScanVCA}). The variational parameters $\matx$ used within \vcaom are the on-site energy of the impurity $\epsilon_f$ and the on-site energies of the uncorrelated cluster sites $\epsilon_s$. The density of states $\rho_f(0)$ displays a pronounced plateau which is related to the existence of a quasi particle peak (Kondo resonance) pinned at the chemical potential. The parameter regions leading to an empty ($-\epsilon_f<0$) or to a doubly occupied ($-\epsilon_f>U$) impurity do not show a pinning of the Kondo resonance at the Fermi energy, as expected. In the half filled region which lies in between, virtual spin fluctuations lead to a pronounced quasi particle peak at the chemical potential. We observe that the result converges with increasing length of the interacting part of the reference system $L$ to the physically expected result. Due to the variational parameters considered, the deviations of the results as a function of $L$ are rather small as compared to CPT where the results change significantly with increasing size of the reference system (see \fig{fig:efScanCPT}). We expect CPT calculations in the empty or doubly occupied regions to converge rather fast (within a few sites) while calculations in the Kondo regime, and particularly in the crossover region, may fully converge only at very large (i.e. exponentially) sizes of the reference system~\cite{hand_spin_2006}. This is inferred from the spin-spin correlation function in the cluster which is observed to decay sufficiently fast outside the Kondo plateau (i.e. it is effectively zero at the boundary of the cluster) but shows long range correlations inside the plateau. The \vcasc results are obtained with one variational parameter $\matx=\{\epsilon_f\}$. The reason for not using $\matx=\{\epsilon_f,\epsilon_s\}$ again is that the result is almost the same as the one obtained with \vcaom (see \fig{fig:efScanVCA}). However in some (small) parameter regions the numerical evaluation becomes difficult. The \vcasc data shown in \fig{fig:efScanSCN} shows a clear improvement as compared to CPT but does not reach the quality of the \vcaom result in terms of convergence in system size.\\
The Friedel sum rule (FSR)~\cite{langer_friedel_1961,langreth_friedel_1966,hewson_kondo_1997} provides an exact relation between the extra states induced below the Fermi energy by a scattering center and the scattering phase shift. It also holds true for interacting systems. This gives a relation between the f-orbital occupation $\langle n^f\rangle$, and the density of states at the Fermi energy:
\begin{align} 
  \rho_{f}(0) &= \frac{1}{\pi\Delta}\; \sin^2\left(\frac{\pi \langle n^f\rangle}{2}\right)\;\mbox{.}
\label{eq:FSR}
\end{align}
In our case the mean occupation in the Kondo regime is $\langle n^f\rangle\approx1$. Since both, the occupancy of the f-orbital and the magnitude of the local density of states at the Fermi energy, can be evaluated independently, we can check the validity of the Friedel sum rule in our approximation. Results are shown in \fig{fig:efScanVCA} applied to the $L=8$ site \vcaom data. The \vcaom results fulfill the Friedel sum rule almost in the whole Kondo region. At the crossover to an empty or doubly occupied impurity the Friedel sum rule is not fulfilled exactly any more but approximated very well. Further outside the agreement is again excellent. The variational parameters of VCA are crucial to fulfill the Friedel sum rule as can be seen from a CPT calculation (\fig{fig:efScanCPT}) which violates it in all parameter regions. It appears that \vcaom with variational parameters $\matx=\{\epsilon_f,\epsilon_s\}$ naturally drives the system to fulfill this condition. The \vcasc result (\fig{fig:efScanSCN}) violates the sum rule too. This is not a feature of \vcasc in general but has to do rather with the choice of variational parameters, which was just $\matx=\{\epsilon_f\}$ in this case. The \vcasc result for two variational parameters $\matx=\{\epsilon_f,\epsilon_s\}$ looks qualitatively like the respective \vcaom result.\\
Scanning the interaction strength $U$ at fixed impurity on-site energy $\epsilon_f$ confirms the presence of the Kondo behavior. Shown in \fig{fig:UScan} are results obtained with \vcaom using the same variational parameters $\matx=\{\epsilon_f,\epsilon_s\}$ as above. In the weakly correlated part ($ U/\Delta \lesssim 5$) the density of states at the chemical potential is low. The intermediate region ($5 \lesssim U/\Delta \lesssim 15$) signals the crossover to the Kondo regime. For larger $U$ the Kondo regime is reached with an impurity occupation of $\langle n^f\rangle\approx 1$, which may be inferred from the Friedel sum rule. In the inset of the figure, the CPT results for the same lengths of the interacting part of the reference system $L$ are shown. The CPT results are by far not converged for the interacting cluster sizes considered here. This emphasizes the importance of the variational parameters.\\
Our results in \fig{fig:efScanVCA} and \fig{fig:UScan} agree very well with those of calculations based on X-operator technique exercised by Lobo \etal~\cite{lobo_atomic_2010}. In their work a strong coupling perturbation theory is applied starting from the Anderson molecule as a basis and using the Friedel sum rule as a condition to fix the position of an infinitely narrow conduction band.\\
Analytic considerations (see \app~\ref{sec:AppendixFSR}) allow insight into the behavior of the Friedel sum rule in ED, CPT and VCA. There it is shown that ED always has to violate the Friedel sum rule while CPT always fulfills it in the particle-hole symmetric case. This comes about in the first place because the height of the Kondo resonance at $\omega=0$ does not depend on the self-energy. A pinning of the Kondo resonance however can only be achieved via the improved self-energy contributions obtained within VCA.\\
The results of this section clearly show that VCA is able to capture the basic physics of the SIAM in every parameter region. The improvement obtained by going over from CPT to VCA is crucial to fulfill exact analytic relations. Moreover we have shown that CPT/VCA is clearly superior to ED calculations. VCA is, even at small $L$, capable of fulfilling the FSR also away from particle-hole symmetry due to a pinning of the Kondo resonance at the Fermi energy. This pinning can be attributed to the better approximation of the self-energy of VCA with respect to CPT.\\

\subsection{\label{sec:Phasediagram}Crossover diagram}
\begin{figure}
        \centering
        \includegraphics[width=0.48\textwidth]{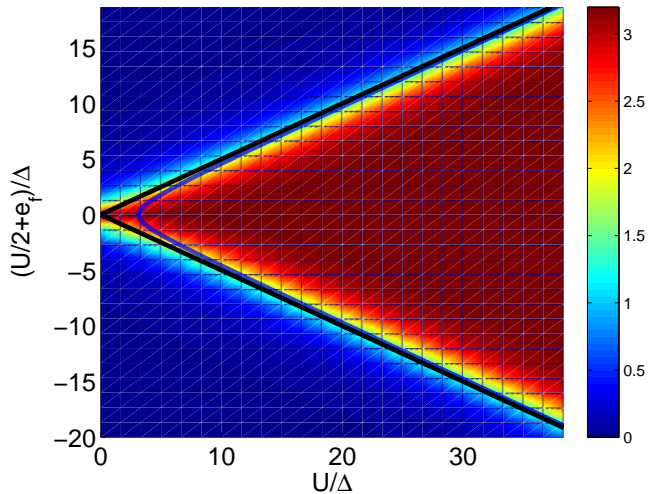}
        \caption{(Color online) In this plot a ``phase diagram'' of the SIAM is shown. The quantity on the z-axis is the density of states in the impurity $\rho_f$ at $\omega=0$ (i.e. the height of the Kondo resonance). The results are obtained with \vcaom for a set of variational parameters $\matx=\{\epsilon_f,\epsilon_s\}$, $L= 6$ and $0^+=10^{-6}$. The black line indicates the different regions obtained from an atomic limit calculation. In the right cone local moments are to be expected. While in the upper region the impurity is expected to be empty and in the lower half to be doubly occupied. The blue curve shows the onset of a spurious magnetic state as obtained by a mean field treatment (see text).}
        \label{fig:PhaseDiagramRho}
\end{figure}

\begin{figure}
        \centering
        \includegraphics[width=0.48\textwidth]{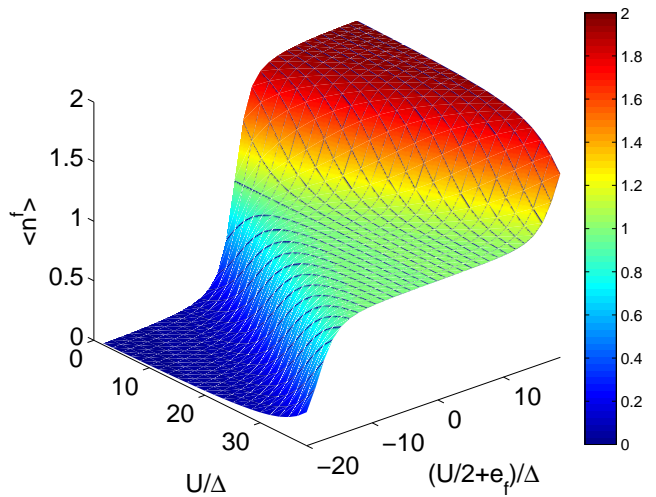}
        \caption{(Color online) Average particle density in the impurity $\langle n^f\rangle$ as a function of $U$ and $\epsilon_f$ for the same parameters as in \fig{fig:PhaseDiagramRho}.}
        \label{fig:PhaseDiagramN}
\end{figure}

To delve into the CPT/VCA results for the whole parameter range of the SIAM a ``phase diagram'' is presented in this section. This should be understood to be a mere scan of the parameters $U$ and $\epsilon_f$ because the model does not undergo a phase transition. The density of states of the impurity at the chemical potential $\rho_f(0)$ is shown in \fig{fig:PhaseDiagramRho} in a density plot. This figure essentially shows the height of the Kondo resonance as a function of interaction strength and on-site energy of the impurity. The different regimes of the SIAM, as obtained by an atomic limit calculation, are indicated as black lines. These lines divide the physics into regions where the impurity is doubly, singly or not occupied. In the singly occupied region ($\frac{U}{2}>|\epsilon_f + \frac{U}{2}|$) local moments and their screening are expected to appear. This region, which bestrides the cone enclosed by black lines, is the region where Kondo physics may take place within this approximation. The parameter regions where the impurity is empty or doubly occupied lie above and below this cone. More sophisticated methods will lead to a smearing out of the border of these regions and introduce a crossover area with competing effects. A boundary expected between a single resonance and a spurious local moment behavior where the single resonance is split into two for spin up and spin down respectively is obtained by mean field theory~\cite{coleman_local_2002}. In the mean field approach the density-density interaction of the impurity Hamiltonian is replaced by a spin dependent density exposed to the mean contribution of the other spins' density. The mean field boundary is then obtained by replacing the mean field parameters $\langle n_{\uparrow}^f\rangle$ and $\langle n_{\downarrow}^f\rangle$ by the particle number $n$ and magnetization $m$. Setting the magnetization to $m=0^+$ in the self consistent equations yields the implicit result $U_c= \pi \Delta (1 +\cot(\pi \frac{\langle n_f\rangle}{2})^2)$ and $\epsilon_{f,c} = \Delta (\cot(\pi \frac{\langle n_f\rangle}{2})+\frac{\pi}{2}\,(1-\langle n_f\rangle) \,(1+\cot(\pi \frac{\langle n_f\rangle}{2})^2))$. The plot shows that the Kondo plateau is reproduced very well by VCA$_\Omega$. The results appear almost converged for lengths $L\approx 6$ of the interacting part of the reference system. Increasing $L$ yields better results in the crossover region. Results obtained by means of CPT do not reproduce the Kondo plateau very well for small $L$.\\ 
The average impurity occupation for the same parameter region is shown in \fig{fig:PhaseDiagramN}. The result obtained with VCA$_\Omega$ clearly shows the Kondo plateau where the impurity is singly occupied. The parameter regions of a doubly occupied or empty impurity lead to a density of states in the impurity which is zero at the chemical potential (compare to \fig{fig:PhaseDiagramRho}).\\
The results of this section have been obtained using \vcaom with variational parameters $\matx=\{\epsilon_f,\epsilon_s\}$. It should be noted that using only $\matx=\{\epsilon_f\}$ already yields good results. As mentioned in \se~\ref{sec:Efscan}, CPT needs very large values $L$ to yield the same quality of the results as VCA does with much smaller values of $L$.\\

\subsection{\label{sec:LowEnergy}Low energy properties, Kondo Temperature}
In this section we examine the low energy properties of the symmetric SIAM. In the strong coupling limit a single scale, the Kondo temperature $T_K$, governs the low energy physics~\cite{hewson_kondo_1997}.\\
The Kondo temperature $T_K$ is known from Bethe Ansatz results for the particle-hole symmetric SIAM~\cite{fateev_exact_1981, kawakami_exact_1981}
\begin{align}
T_K &= \sqrt{\frac{\Delta U}{2}}\,e^{-\gamma\frac{\pi}{8 \Delta}\,U}\;\mbox{,   }\;\gamma=1\;\mbox{.}
 \label{eq:Tk}
\end{align}
This scale, which is inversely proportional to the spin-flip rate of the impurity, divides the physics of the SIAM into two regions: A local moment behavior of the impurity, where the spin is free, and a low temperature region where the local spin and the conduction electrons become entangled and form a singlet state~\cite{fulde_electron_2003}.\\
Quantities which depend inversely on $T_K$ are the effective mass $m^*$ and the static spin susceptibility $\chi_m$. The Kondo temperature may furthermore be extracted from the width or weight of the Kondo resonance in the local density of states of the impurity f-orbital. We investigate and compare the results for the scale $T_K$ obtained from the direct determination of $T_K$ (from the FWHM and the spectral weight of the Kondo resonance) and the inverse quantities $m^*$ and $\chi_m$. We find that the results of all four measurements turn out to yield the correct qualitative behavior in \vcaomNOBLANK. However, in a region where the dependence of $T_K$ is exponentially dependent on the interaction strength $U$ the exponential prefactor is not predicted correctly. Therefore we introduce a scaling factor $\gamma$ (\eq{eq:Tk}) which turns out to be the same for all four ways of determining $T_K$. In particular this factor is independent of the set of model parameters used. The scaling factor may be calculated semi-analytically for a reference system consisting of a two site interacting cluster and the semi-infinite environment within \vcaom and \vcasc ($\matx=\{V\}$). The calculation for \vcaom leads to an integral expression for the stationary point of the grand potential $\Omega$ with respect to $\Delta V$ from which the optimal $\Delta V$ can be obtained numerically (see \app~\ref{sec:AppendixTwoSiteAnalytical}). The Kondo scale may be determined from the so obtained values of $V'(U)=\Delta V(U) + V$ by
\begin{align}
T_K(U) \propto \left(\frac{V'(U)}{U}\right)^2\;\mbox{.}
\label{eq:TkTwoSite}
\end{align}
This leads to a perfect exponential behavior as defined in \eq{eq:Tk} with
\begin{align*}
\gamma = 0.6511\;\mbox{.}
\end{align*}
The issue of obtaining an exponential scale but not the correct exponent for the functional dependence on $U$ is common to various approximate methods (for example variational wave functions where the issue was cured by introducing an extended Ansatz by Sch\"onhammer~\cite{schoenhammer_variational_1976}, saddle-point approximations of a functional integral approach~\cite{kotliar_new_1986} or FRG~\cite{andergassen_gentle_2006}). A faint analogy may be drawn here to Gutzwiller approximation, where an exponential energy scale in $U$ arises by a renormalized hybridization parameter $V$~\cite{Tsunetsugu1997}, which is also the case for \vcaomNOBLANK.\\
The self consistent calculation for \vcasc also leads to an integral expression for the determination of $\Delta V$. This expression is obtained by requiring the expectation values of the hopping from the impurity f-orbital to the neighboring site in the reference system to be the same as the expectation value in the physical system. This procedure does not yield an exponential scale in $U$. The optimal cluster parameter $V'$ shows spurious behavior as a function of $U$.
We conclude that \vcasc with $\matx=\{V\}$ cannot reproduce the low energy properties of the SIAM even qualitatively, while \vcaom yields the correct behavior apart from an exponential factor.\\
\subsubsection{\label{sec:EffectiveMass}Effective mass - quasi particle renormalization}
\begin{figure}
        \centering
        \includegraphics[width=0.48\textwidth]{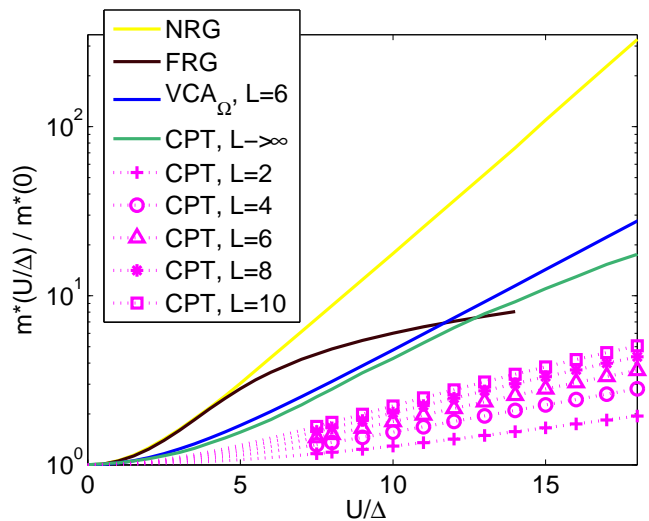}
        \caption{(Color online) Effective mass $m^*$ of the Kondo resonance \eq{eq:meff} as a function of interaction strength $U$. We plot CPT results for lengths of the interacting part of the reference system $L=2,4,6,8 \text{ and } 10$ (magenta), the to $L\rightarrow\infty$ extrapolated CPT result (olive), as well as \vcaom results (blue). The data points for the CPT result in the low $U$ region are not shown to avoid messing up the plot. The variational parameter used for the \vcaom result was $\matx=\{V\}$. The \vcaom data was obtained for $L= 6$. For CPT as well as \vcaomNOBLANK, we used a numerical broadening of $0^+=10^{-6}$. For comparison the results obtained by NRG (yellow) and FRG (dark brown) are shown~\cite{karrasch_finite-frequency_2008}.} 
        \label{fig:mStar}
\end{figure}

The effective mass $m^*$ is defined as the quasi particle renormalization~\cite{karrasch_finite-frequency_2008}
\begin{align}
\nonumber \frac{m^*(U)}{m^*(0)} &= 1-\frac{d[\Im{\text{m}}\,\Sigma^\sigma_{ff}(i\omega,U)]}{d\,\omega}\bigg|_{\omega=0^+} \\
\nonumber &= \frac{d[\Im{\text{m}}\,\GF^\sigma_{ff}(i\omega,U)]}{d\,\omega}\bigg|_{\omega=0^+}\,\times\\
&\left(\frac{d[\Im{\text{m}}\,\GF^\sigma_{ff}(i\omega,0)]}{d\,\omega}\bigg|_{\omega=0^+}\right)^ {-1}\;\mbox{,}
 \label{eq:meff}
\end{align}
where we introduced the dependence on the interaction strength $U$ explicitly. In the Kondo regime, this quantity becomes inversely proportional to the Kondo temperature.\\
We want to answer the question whether the Kondo scale is approximately captured by CPT/VCA or not. Therefore we compare the functional form and the exponent obtained from the effective mass and the analytic result for $T_K$, \eq{eq:Tk}. The result for the effective mass obtained within \vcaom is shown in \fig{fig:mStar}. The variational parameter used was $\matx=\{V\}$. The functional form is reproduced well by \vcaom (i.e. it starts out quadratically and goes over to an exponential behavior in the Kondo region). However the exponent $(\frac{\pi}{8 \Delta})$ is not reproduced correctly. \vcaom yields a lower exponent of $\approx(\gamma\,\frac{\pi}{8 \Delta})$. The factor $\gamma$ is defined in \app~\ref{sec:AppendixTwoSiteAnalytical}, determined from a semi-analytical calculation of $T_K$ within \vcaomNOBLANK. This additional factor is the same for all initial parameters (within the Kondo regime), it is particularly independent of $\Delta$. If plotted over a scaled $U$ axis, $U'=\frac{1}{\gamma}U$, the \vcaom result would lie on top of the NRG data. However using larger sizes of the interacting part of the reference system does not lead to much better results regarding $\gamma$. It is to be expected, that a significant improvement can only be obtained using exponentially large $L$. The CPT result shows a very different convergence behavior in $L$ which is rather slow.\\
An attempt was made to extrapolate the CPT data to $L\rightarrow\infty$ by a simple $\frac{1}{L}$ scaling. It is interesting to observe that this extrapolated curve coincides nicely with the VCA result ($L=6$) in the low $U$ region. However, we expect this extrapolation based on small $L$ to be insufficient to capture the exponential scaling of the data in $L$. Note that since in CPT the self-energy is taken from the cluster, the CPT results for the effective mass coincide with ED results for systems of size $L$.\\

\subsubsection{\label{sec:WeightTemperature}Kondo spectral weight and half width}
\begin{figure}
        \centering
        \includegraphics[width=0.48\textwidth]{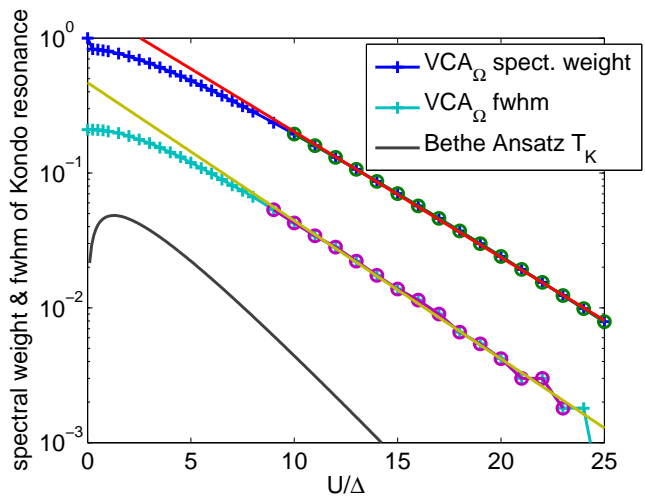}
        \caption{(Color online) \vcaom results for the spectral weight (blue) and  full width at half maximum (olive) of the Kondo resonance as a function of interaction strength $U$. The variational parameter used was $\matx=\{V\}$. A length of the interacting part of the reference system of $L= 6$ sites and a numerical broadening of $0^+=10^{-6}$ were used for this calculation. Data points marked with a circle were used for the fit of the exponential function in the Kondo region. The black line shows the Kondo temperature $T_K$ as obtained by Bethe Ansatz calculations \eq{eq:Tk}.}
        \label{fig:TkSpectWeightFWHM}
\end{figure}

Since the height of the Kondo resonance is fixed by the Friedel sum rule \eq{eq:FSR} the width and the weight (area) of the peak are proportional to the Kondo temperature $T_K$. Obtaining the spectral weight or FWHM of the Kondo resonance from the spectrum introduces a large uncertainty. Nevertheless we made an attempt, to get an idea of the behavior  of $T_K$. We fixed the spectral weight by the first minimum to the left and to the right of the central peak (see also \tcite{hand_spin_2006}). In general the effective mass and static spin susceptibility will yield more reliable results but it is instructive to compare these four ways of determining $T_K$.\\
Shown in \fig{fig:TkSpectWeightFWHM} is the evolution of the spectral weight and the FWHM of the Kondo resonance with increasing interaction strength $U$. The data was acquired using VCA$_{\Omega}$ with a variational parameter $\matx=\{V\}$ for the particle-hole symmetric SIAM. Within the uncertainty, the same exponential behavior for the Kondo temperature $T_K$ is obtained as by calculating the effective mass in \vcaomNOBLANK.\\

\subsubsection{\label{sec:StaticSpinSusceptibility}Static spin susceptibility}
\begin{figure}
        \centering
        \includegraphics[width=0.48\textwidth]{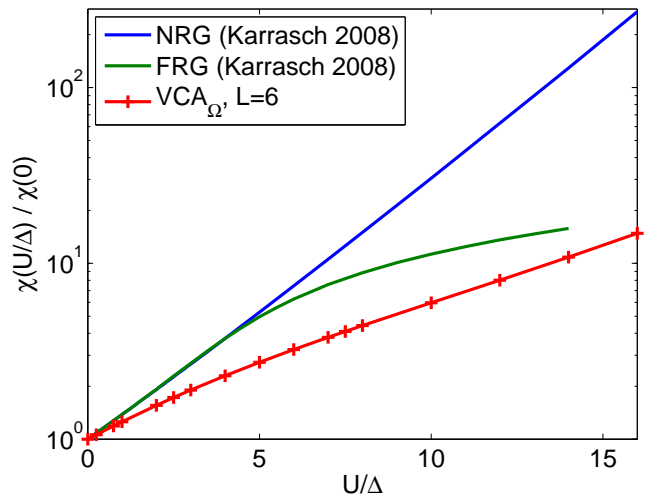}
        \caption{(Color online) The static spin susceptibility $\chi_m$ \eq{eq:chim} is shown as a function of interaction strength $U$. The variational parameter used was $\matx=\{V\}$. The data was obtained for $L= 6$ sites and $0^+=10^{-6}$. For comparison the results obtained by NRG (blue) and FRG (green) are shown~\cite{karrasch_finite-frequency_2008}.}
        \label{fig:chiM}
\end{figure}

The static spin susceptibility $\chi_m$ is given by the linear response to an applied magnetic field $B$ in z direction
\begin{align}
\chi_m(U) &= -\frac{d\left(\left\langle n^f_{\uparrow}\right\rangle-\left\langle n^f_{\downarrow}\right\rangle\right)}{d\,B}\bigg|_{B=0}\;\mbox{.}
 \label{eq:chim}
\end{align}
In the Kondo regime this quantity too becomes inversely proportional to the Kondo temperature. For the calculations in this section we introduce an additional spin dependent term in the impurity Hamiltonian (\eq{eq:Himpurity})
\begin{align}
\hat{\mathcal{H}}_{\text{magnetic}} &= \sum\limits_{\sigma}  \sigma\frac{B}{2}\, f_{\sigma}^\dagger \, f_{\sigma}^\nag \mbox{.}
 \label{eq:Hmagnetic}
\end{align}
The static spin susceptibility $\chi_m$ as obtained with \vcaom is shown in \fig{fig:chiM}. The variational parameter used was $\matx=\{V\}$. As a reference the results of NRG and FRG~\cite{karrasch_finite-frequency_2008} are shown. The behavior of the VCA result is good for small interaction strength $U$. The VCA result shown for $L=6$ appears already converged while the CPT result would require much larger systems.\\
We would like to highlight that \vcaom reproduces an energy scale $T_K$. Results from direct calculation of $T_K$, calculation of the effective mass $m^*$ and the static spin susceptibility $\chi_m$ yield the correct functional form but not the right exponent.\\

\subsection{\label{sec:CTQMC}Benchmarking CPT/VCA against continuous time Quantum Monte Carlo}
\begin{figure}
        \centering
        \includegraphics[width=0.48\textwidth]{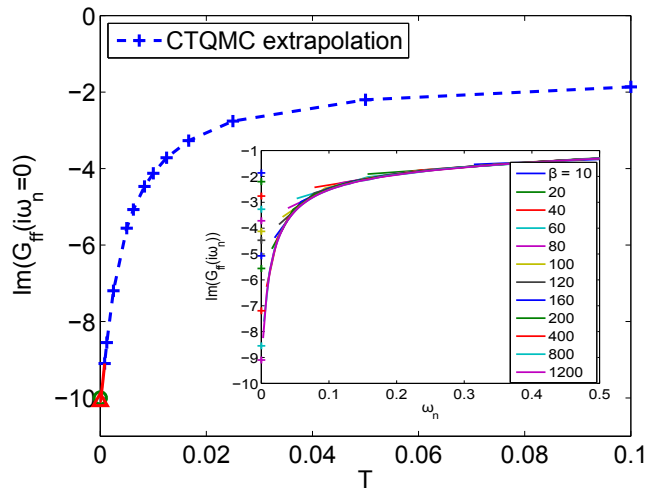}
        \caption{(Color online) CT-QMC result for the imaginary part of the impurity Green's function extrapolated to $i\omega_n=0$. An extrapolation to zero temperature is attempted, which yields a good agreement with the result predicted by the Friedel sum rule (green circle)  within the uncertainty (red triangle). The  inset shows the imaginary part of the impurity Green's function for various $\beta$ (see legend) and the extrapolated points at $i\omega_n=0$.}
        \label{fig:CTQMCextrapolation}
\end{figure}

\begin{figure}
        \centering
        \includegraphics[width=0.48\textwidth]{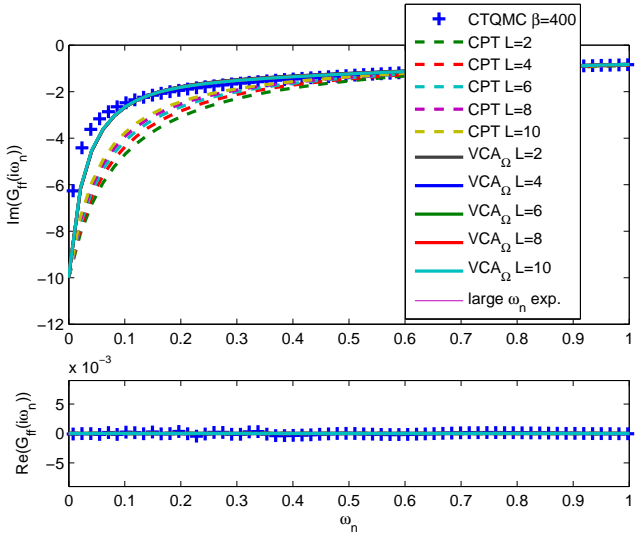}
        \caption{(Color online) Comparison of the Matsubara impurity Green's function $G_{ff}(i\omega_n)$ obtained by CT-QMC ($\beta=400$), CPT and VCA$_{\Omega}$. The CPT/VCA results were obtained for $L=2,4,6,8 \text{ and } 10$. The real part shown in the lower part of the figure is zero. The Friedel sum rule prediction of $\Im{\text{m}(G_{ff}(i\omega_n=0))}=-10$ is fulfilled by all methods. The legend of this figure serves as well as legend for \fig{fig:SIGMA_LARGE_IWCPT} and \fig{fig:CTQMCvsCPT_SIGMA}. That is why the last entry (large $i\omega_n$ exp. (see \eq{eq:SigmaLargeIW})) is displayed in the legend but is missing in the graph of this figure.}
        \label{fig:CTQMCvsCPT_G}
\end{figure}

\begin{figure}
        \centering
        \includegraphics[width=0.48\textwidth]{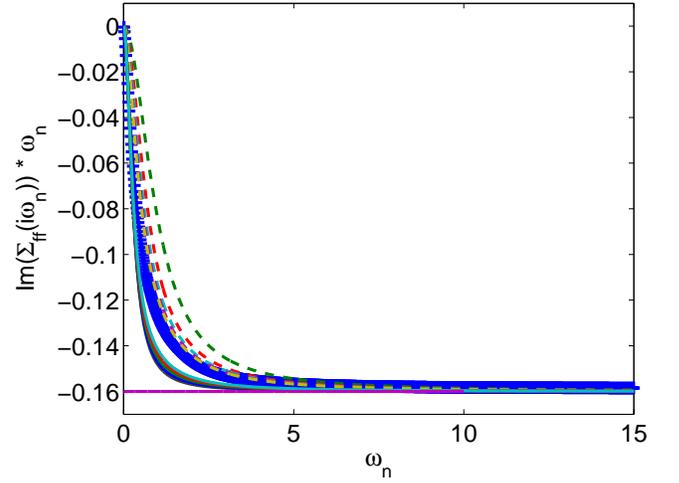}
        \caption{(Color online) Comparison of the self-energy of the impurity $\Sigma_{ff}(i\omega_n)$ times energy $\omega_n$ obtained by CT-QMC ($\beta=400$), CPT and VCA$_{\Omega}$. The CPT/VCA results were obtained for $L=2,4,6,8 \text{ and } 10$. CPT as well as \vcaom become exact for high Matsubara frequencies. An expansion of $\Sigma(i\omega_n)$ for large $i\omega_n$ \eq{eq:SigmaLargeIW} is additionally shown (straight line at $-\left(\frac{U}{2}\right)^2$). The legend for this figure is the same as for \fig{fig:CTQMCvsCPT_G} and is displayed there.}
        \label{fig:SIGMA_LARGE_IWCPT}
\end{figure}

\begin{figure}
        \centering
        \includegraphics[width=0.48\textwidth]{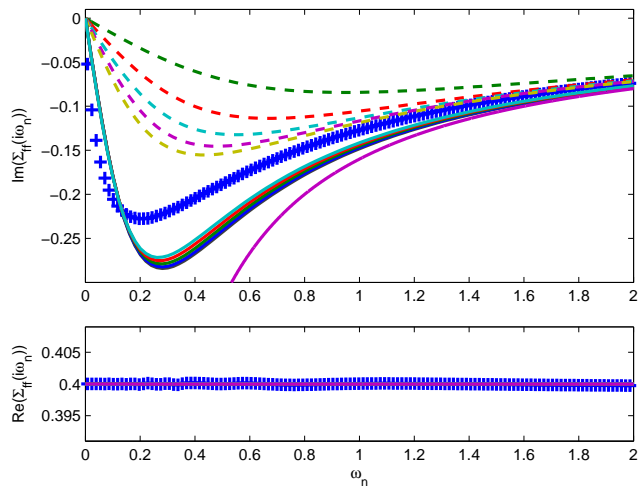}
        \caption{(Color online) Comparison of the imaginary part of the self-energy of the impurity $\Im{\text{m}(\Sigma_{ff}(i\omega_n))}$ obtained by CT-QMC ($\beta=400$), CPT and VCA$_{\Omega}$. The CPT/VCA results were obtained for $L=2,4,6,8 \text{ and } 10$. An expansion of $\Sigma(i\omega_n)$ for large $i\omega_n$ \eq{eq:SigmaLargeIW} is shown in addition (magenta line which diverges at zero). CPT/VCA always reproduces the exact self-energy for high Matsubara frequencies. The legend for this figure is the same as for \fig{fig:CTQMCvsCPT_G} and is displayed there.}
        \label{fig:CTQMCvsCPT_SIGMA}
\end{figure}

In this section we compare CPT/VCA results to QMC data. We obtained the Monte Carlo results using the continuous time Quantum Monte Carlo (CT-QMC) code of the TRIQS~\cite{ferrero_triqs_2011} toolkit and its implementation of the hybridization expansion (CT-HYB)~\cite{werner_hybridization_2006} algorithm using Legendre polynomials~\cite{boehnke_orthogonal_2011}. This method enables access to very low temperatures and is especially suited to obtain low energy properties~\cite{gull_continuous-time_2010}. The CT-QMC data provides statistically exact and reliable results to test our data.\\ 
All CT-QMC calculations were done for a single impurity orbital at $U=0.8$ and $\epsilon_f=-0.4$. We used a semicircular hybridization function with half bandwidth $D=2$ and $V=0.3162$. This setup corresponds to  the same model under investigation here. The value for the interaction strength $U=0.8$ was chosen because of the relatively low expected Kondo temperature of $\beta_K = T_K^{-1} \approx 100$. For all calculations $1.2\e{9}$ MC updates where conducted, with a sweep size of $100$ updates, plus a $10\%$ thermalization period.\\
To ensure that the Kondo resonance is correctly reproduced by CT-QMC we evaluated the Matsubara Green's function for various values of inverse temperature $\beta$. The height of the Kondo resonance is given by the Friedel sum rule \eq{eq:FSR} to be $\Im{\text{m}(G_{ff}(i\omega_n = 0))} =-10$ for the parameters used here ($\Delta=0.1$). To obtain $\Im{\text{m}(G_{ff}(i\omega_n = 0))}$ we extrapolate twice, first in $i\omega_n\rightarrow0$ for each $\beta$, then we use these results and extrapolate to $T\rightarrow0$. The extrapolation to $i\omega_n\rightarrow0$ is done linearly using the first two Matsubara frequencies. The imaginary part of $\GF_{ff}(i\omega_n)$ and the extrapolated value to $i\omega_n\rightarrow0$ are shown in the inset of \fig{fig:CTQMCextrapolation} for $\beta \in [10, 1200]$. Those extrapolated values are plotted as a function of temperature (\fig{fig:CTQMCextrapolation}). These data points are then extrapolated to $T\rightarrow 0$ using a fit by a rational model function. The result clearly shows the onset of the Kondo resonance when the temperature is lowered below the Kondo temperature $T_K$. The extrapolation to $T=0$ shows very good agreement ($\Im{\text{m}(G_{ff}(i\omega_n= 0))} \approx -10.1$) with the result expected from the Friedel sum rule within the uncertainty. It is important to note that the CT-QMC results converge very nicely in $\beta$. Although for higher $\beta$ lower Matsubara frequencies become available, the overall shape of the Green's function does not change significantly.\\ 
Therefore we may compare the $T=0$ CPT/VCA results for the Green's function and self-energy to the CT-QMC data. The Matsubara Green's functions of the impurity f-orbital $G_{ff}(i\omega_n)$ obtained by CT-QMC ($\beta=400$), CPT and VCA  are shown in \fig{fig:CTQMCvsCPT_G}. We use $\beta=400$ as a compromise between low temperatures and still reliable CT-QMC results (within manageable computation time). The $\beta=400$ result was obtained using $65$ Legendre coefficients. A detailed analysis has shown that this number is sufficient to get high frequency moments of the self-energy $\Sigma$ accurately. The VCA$_{\Omega}$ results were obtained with one variational parameter $\matx=\{V\}$ for $U=0.8$, $\Delta=0.1$ and $0^+=10^{-6}$ in the particle-hole symmetric case. For the CPT calculation we used the same parameters. For both methods we considered lengths of the interacting part of the reference system of $L=2,4,6,8 \text{ and } 10$. The VCA result lies near the CT-QMC data but underestimates the slope of the curve at low $i\omega_n$. The VCA result provides a huge improvement upon CPT for the lengths of the interacting part of the reference system shown here. The real part of $G_{ff}(i\omega_n)$ is exactly zero within CPT/VCA as it is supposed to be. Note that the value of $G_{ff}(i\omega_n=0)$ which is fixed by the Friedel sum rule is exactly reproduced within CPT and VCA for the particle-hole symmetric case. 
The same is shown for the self-energy of the impurity f-orbital $\Sigma_{ff}(i\omega_n)$ in \fig{fig:CTQMCvsCPT_SIGMA}. From the imaginary part of $\Sigma_{ff}(i\omega_n)$ one can infer the convergence of the CPT/VCA result with larger length of the interacting part of the reference system $L$. The real part of the self-energy ($\Re{\text{e}(\Sigma_{ff}(i\omega_n)}=\mu=-e_f=\frac{U}{2}=0.4)$ is again exactly reproduced within CPT/VCA.\\
In the following, we discuss the self-energy $\Sigma(i\omega_n)$ for the two interesting cases of very low and very high Matsubara frequency. We start out by conducting an expansion of the self-energy $\Sigma(z)$ for high Matsubara frequencies ($z=i\omega_n \rightarrow \infty$) which shall be outlined here briefly. The self-energy matrix is defined by
\begin{align*}
 \Sigma(z) &=\GF^{-1}_0-\GF^{-1}\\
&= z-\TF-\GF^{-1}\;\mbox{.}
\end{align*}
Here $\TF$ is the one-particle part of the Hamiltonian. In the particle-hole symmetric case considered here it contains all the hoppings as well as the on-site energy of the impurity $\epsilon_f = -\frac{U}{2}$. We conduct a series expansion in powers of $z^{-1}$ of $\Sigma(z)$. Apart from the real constant $\TF_{ii}$ all $z$-dependent terms of $\Sigma_{ii}(z)$ are anti-symmetric in $z$. Therefore even powers in $z^{\pm 2l}\;\mbox{,}\;l>0$ vanish. Expanding the Green's function $\GF(z)$ yields for the self-energy $\Sigma(z)$
\begin{align*}
\Sigma(z) &= -\TF - z \sum\limits_{m=1}^{\infty} (-1)^{m}\,X^m\;\mbox{,}\\
X &= \sum\limits_{n=1}^{\infty} z^{-n}\, C_n\;\mbox{,}\\
(C_n)_{ij} &=  \bra{\Psi_0}a_i\nag(\Delta \hat{\mathcal{H}})^n a_j^\dagger\ket{\Psi_0}\\
&+(-1)^n\,\bra{\Psi_0}a_j^\dagger(\Delta \hat{\mathcal{H}})^n a_i^\nag\ket{\Psi_0}\;\mbox{,}
\end{align*}
where $\Delta \hat{\mathcal{H}} = \hat{\mathcal{H}}-\omega_0$ and $\omega_0$ is the ground-state energy of $\hat{\mathcal{H}}$. Collecting powers of $z$ yields a cumulant-like expansion for the self-energy $\Sigma(z)$
\begin{align*}
 \Sigma(z) &= \sum\limits_{n=1}^{\infty} z^{-n}\, \Sigma_n\;\mbox{, where}\\
 \Sigma_0 &= -T+C_1\;\mbox{, and}\\
 \Sigma_1 &= C_2-C_1^2\;\mbox{.}\\
\end{align*}
Here we consider the zeroth and first order in $z^{-1}$ only and obtain for $\Sigma(i\omega_n)$ 
\begin{align}
\Sigma_{ff}(i\omega_n) = \frac{U}{2} - \frac{i}{\omega}\,\left(\frac{U}{2}\right)^2 + \mathcal{O}\left(\frac{1}{i\omega_n}\right)^3\;\mbox{,}
\label{eq:SigmaLargeIW}
\end{align}
where the self-energy at the impurity f-orbital $\Sigma_{ff}$ is the only non-vanishing matrix element of $\Sigma_{ij}$. This result is plotted as a reference in \fig{fig:CTQMCvsCPT_SIGMA}. Due to the nature of the CPT/VCA approximation these methods always yield the exact self-energy for high Matsubara frequency as shown in \fig{fig:SIGMA_LARGE_IWCPT}.

The low energy properties examined in the previous section depend basically on the slope of the Matsubara Green's function at $(i\omega_n)=0^+$. The results shown in \fig{fig:CTQMCvsCPT_G} and \fig{fig:CTQMCvsCPT_SIGMA} show that this slope is underestimated by CPT/VCA in comparison to CT-QMC, at least at the small lengths of the interacting part of the reference system available.\\
The above results suggest a possible application of VCA as an impurity solver for zero temperature DMFT. The results would not suffer from a bath truncation error as in exact diagonalization based DMFT. A big advantage would be the low demand on computational power of VCA as well as the approximate reproduction of the main features of the local density of states (i.e. Kondo resonance and high energy incoherent part of the spectrum).\\

\subsection{\label{sec:Bx}Introducing a symmetry breaking field}
\begin{figure}
        \centering
        \includegraphics[width=0.48\textwidth]{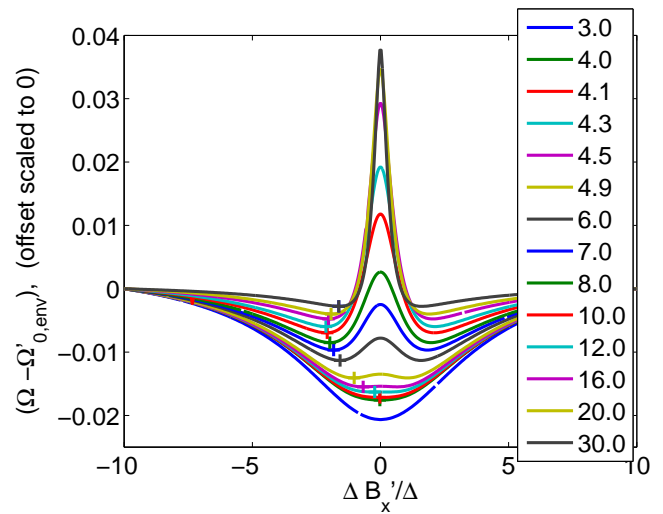}
        \caption{(Color online) Grand potential $\Omega-\Omega'_{0,env}$ (\eq{eq:omegaResult}) as a function of the interaction strength $U/\Delta$ (see legend). The data was obtained by studying a $L=4$ site interacting cluster coupled to a semi-infinite lead. The numerical broadening used was $0^+=10^{-6}$. The crosses indicate the respective minimum of the grand potential. There exists a critical $U_c/\Delta\approx4.3$ above which a finite $B_x'$ is preferred by the system.}
        \label{fig:Bx}
\end{figure}

\begin{figure}
        \centering
        \includegraphics[width=0.48\textwidth]{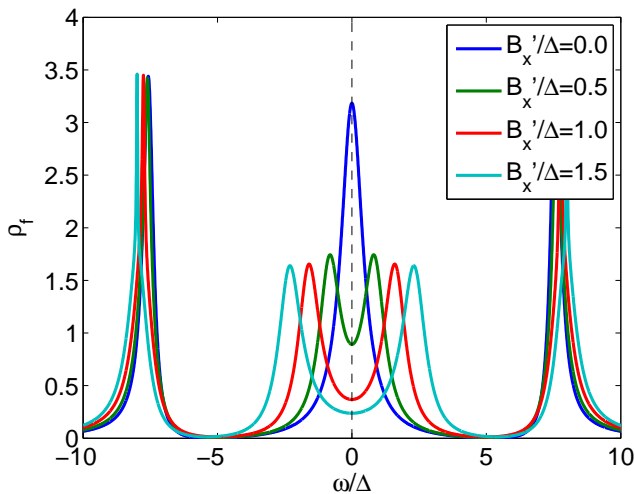}
        \caption{(Color online) The splitting of the Kondo resonance caused by an applied magnetic field in x direction is shown for different values of the auxiliary field $B_x'$. The plots were obtained using VCA (i.e. the physical field $B_x$ is always zero). Instead of taking the parameter $B_x'$ at the stationary point of the grand potential (this value would be $B_x'/\Delta\approx1.9$ for the parameters used) we explicitly plug in a fixed value for $B_x'$. The length of the interacting part of the reference system used was $L=6$ for the model parameters $U/\Delta=12$. The numerical broadening used was $0^+=10^{-6}$.}
        \label{fig:BxSpectra}
\end{figure}

We explore the possibility to improve the VCA results achieved by varying the internal single-particle parameters of the model by introducing a symmetry breaking ``spin flip field'' at the impurity f-orbital. The term added to the impurity Hamiltonian \eq{eq:Himpurity}
\begin{align}
\hat{\mathcal{H}}_{\text{flip}} &= B_x \, \left( f_{\uparrow}^\dagger \, f_{\downarrow}^\nag + f_{\downarrow}^\dagger \, f_{\uparrow}^\nag\right) \;\mbox{,}
 \label{eq:HspinFlip}
\end{align}
explicitly breaks the conservation of spin in the cluster solution. We are interested in the model with a physical parameter $B_x=0$ so this variable may only attain a finite value as a variational parameter $B_x'$ in the reference system. We investigate the particle-hole symmetric model at $V=0.3162$ and $t=1$. Our findings indicate that any finite value of $B_x'$ splits the Kondo resonance and has thus to be discarded on physical grounds for the system under investigation.\\
While this prevents the application of this field to improve the VCA results, it gives very nice insight in the physics of the SIAM as described by CPT/VCA. We find that a critical interaction strength $U_c$ depending on the length of the interacting part of the reference system exists which separates solutions which would prefer a finite $B_x'$ from those which would prefer $B_x'=0$. The critical interaction strength for $L=4$ is given by $U_c/\Delta\approx4.3$. The grand potential $\Omega-\Omega'_{0,env}$ is plotted for various interaction strengths $U$ in \fig{fig:Bx}. For an analogous calculation for $L=6$ site interacting clusters a value of $U_c/\Delta\approx4.1$ is achieved. The mean field result would yield a critical interaction strength $U_c/\Delta=\pi$ for the parameters used here. We interpret this value as a signature of the onset of local moment behavior. The values for $U_c$ are of course not to be taken literally, they depend very much on the finite size of the cluster under investigation.\\
The splitting of the Kondo resonance caused by a non-zero variational field $B_x'$ is shown in \fig{fig:BxSpectra}. The value of $U/\Delta=12$ used for this calculation lies in the region above $U_c$ where the system prefers a nonzero field $B_x'$.\\

\section{\label{sec:conclusion}Conclusions}
In this work we have applied the variational cluster approach (VCA) to the single impurity Anderson model. We devised a cluster tiling applicable to this non-translationally invariant model which leads to a cluster with a discrete spectrum and an environment having a continuous spectrum. We have derived an expression for the change of the grand potential originating from the coupling of the impurity to the semi-infinite bath. \\
We have compared results for the single-particle dynamics to data obtained by exact diagonalization and cluster perturbation theory (CPT). We found that the variational extension made by the VCA is vital for a good reproduction of the expected behavior of the SIAM. The CPT/VCA spectra both yield a Kondo resonance in the impurity density of states with the correct height as predicted by the Friedel sum rule. A close look at the Kondo resonance shows that the VCA is able to reproduce the resonance and the functional behavior for the Kondo temperature in a remarkable way. The Kondo temperature is expected to show exponential behavior in interaction strength in the Kondo regime. VCA yielding an exponential behavior however tends to underestimate the exponent. Comparison of dynamic quantities to continuous time Quantum Monte Carlo solidifies the origins of this behavior. The high energy incoherent part of the spectrum shows strong finite size effects within CPT which are partly removed by virtue of the VCA. VCA furthermore reproduces the expected position and width of the high energy part of the spectrum. For the asymmetric model, the Friedel sum rule is fulfilled in all parameter regions implying that the Kondo resonance is pinned at the chemical potential in the Kondo region. In addition a self consistent formulation of the VCA, previously introduced in the context of non equilibrium problems~\cite{knap_nonequilibrium_2011}, was explored. Results obtained by the self consistent approach show agreement with results obtained by VCA based on the grand potential for the density of states of the impurity f-orbital. Thereby the positions of the spectral features agree very well with the traditional VCA result, while the spectral weight distribution may deviate especially for small values of interaction strength. Comparison to results obtained from Bethe Ansatz, renormalization group approaches and data obtained from X-Operator based calculations show reasonable agreement for all quantities investigated. \\
In conclusion, while there are certainly more accurate methods to deal with a single quantum impurity model~\cite{affleck_quantum_2008}, especially at low energies, our work shows that VCA is a flexible and versatile method which provides reasonably accurate results with modest computational resources. Here, the VCA self consistency condition proves to be crucial. This allows to obtain the same accuracy that CPT would provide with a much larger, inaccessible, interacting cluster size. One of the advantages is the flexibility of the method, i.e. it is straightforward to extend it to many impurities, non equilibrium problems~\cite{knap_nonequilibrium_2011}, etc. In the spirit of NRG~\cite{bulla_numerical_2008} one could improve on the present results by carrying out an appropriate unitary transformation on the bath such that the bath hoppings decay with increasing distance from the impurity. In this work, we do not include such an improvement and choose a constant hopping sequence (\eq{eq:Hconduction}), since our goal is to benchmark VCA/CPT only. Nevertheless, a hybrid approach combining NRG and VCA would be an interesting extension of the present work.\\

\begin{acknowledgments}
We are grateful to R. Peters for providing his DMRG and NRG data shown in \fig{fig:spectraPHsym} and to C. Karrasch for providing his FRG and NRG data shown in \fig{fig:spektralInset}, \fig{fig:mStar} and \fig{fig:chiM}. MN wants to thank K. Sch\"onhammer, T. Pruschke, M. Knap and P. Dargel for fruitful discussions. We thank O. Parcollet and M. Ferrero for important, critical feedback concerning the convergence of our CT-QMC data. We made use of the CT-QMC code of the TRIQS~\cite{ferrero_triqs_2011} toolkit. This work is partly supported by the Austrian Science Fund (FWF) P18551-N16. MN acknowledges financial support by the F\"orderungsstipendium of the TU Graz.\\
\end{acknowledgments}

\appendix
\section{\label{sec:AppendixOmeganew}Grand potential}
Here we outline the proof of \eq{eq:aux3}. We start out from \eq{eq:aux0}, i.e.
\begin{align*}
\Delta \Omega &= - \Tr\ln \bigg(\uu - \TF \gf\bigg)\;.
\end{align*}
Taylor expansion yields
\begin{align}\label{eq:aux1}
\Delta \Omega
&=  \sum_{n=1}^{\infty} \frac{1}{n}\;
\left\{
\Tr \bigg[\big(\TF\gf\big)^{n}\bigg]_{cc}
+\Tr \bigg[\big(\TF\gf\big)^{n}\bigg]_{ee}\right\}\;\mbox{.}
\end{align}
Due to $\TF_{ee}=0$ each term  $\gf_{ee}$ in the first trace occurs only in the form
\begin{align*}
\tilde \gf_{cc} &:= \TF_{ce} \gf_{ee} \TF_{ec}\;.
\end{align*}
The expressions in the second part of \eq{eq:aux1} can be modified by a cyclic permutation of the factors in the argument of the trace 
\begin{align*}
\Tr(\TF_{ec} \gf_{cc} \cdots \TF_{ce} \gf_{ee}) &=\Tr( \gf_{cc} \cdots \TF_{ce} \gf_{ee} \TF_{ec})\;,
\end{align*}
in which $\gf_{ee}$ again  only occurs in the dressed form $\tilde \gf_{cc}$. If we replace all occurrences of $\TF_{ce} \gf_{ee} \TF_{ec}$ by $\tilde \gf_{cc}$ then there are no matrices $\TF_{ce}$ and $\TF_{ec}$ respectively, left. Hence we can as well introduce in \eq{eq:aux0} the following replacements
\begin{align*}
\gf_{ee} &\to\tilde \gf_{cc}\;,
\TF_{ce} \to \uu_{cc}\;,
\TF_{ec} \to \uu_{cc}\;,
\uu \to \uu_{cc}\;,
 \end{align*}
as it leads in the series expansion to the same expressions. The argument in \eq{eq:aux0} then assumes the form
\begin{align*}
\uu - \TF\gf &= \uu -
\begin{pmatrix}
 \TF_{cc} & \uu_{cc}\\
 \uu_{cc} & 0_{cc}
\end{pmatrix}
\begin{pmatrix}
\gf_{cc}&0_{cc}\\
 0_{cc}&\tilde \gf_{cc}
\end{pmatrix}
=
\begin{pmatrix}
b & -\tilde \gf_{cc}\\
- \gf_{cc} & \uu
\end{pmatrix}\;,
\end{align*}
with the abbreviation $b:=\uu -  \TF_{cc} \gf_{cc}$. Prompted by the Schur complement, the matrix can be factorized into upper and lower triangular block matrices
\begin{align*}
\begin{pmatrix}
b&-\tilde \gf_{cc}\\
-\gf_{cc} &\uu
\end{pmatrix} 
&=
\begin{pmatrix}
 b & 0\\
 -\gf_{cc} & \uu - \gf_{cc} b^{-1} \tilde \gf_{cc}
\end{pmatrix}
\begin{pmatrix}
 \uu & -b^{-1}   \tilde \gf_{cc}\\
  0 &\uu
\end{pmatrix}\;,
\end{align*}
such that the determinant is easily computed, since the determinant of the second matrix is $1$ and the determinant of the first matrix is simply the product of the determinants of the diagonal blocks, resulting in
\begin{align*}
\det
\begin{pmatrix}
b&-\tilde \gf_{cc}\\
-\gf_{cc} &\uu
\end{pmatrix} 
&=
\det(b) \det\big(\uu - b^{-1} \tilde \gf_{cc} \gf_{cc}\big)\\
&=\det
\bigg(
b - \tilde \gf_{cc} \gf_{cc}
\bigg) \\
&= 
\det
\bigg(
\uu - (\TF_{cc} + \TF_{ce}\gf_{ee}\TF_{ec}) \gf_{cc}
\bigg)\;.
\end{align*}
The final result for \eq{eq:aux0} reads
\begin{align*}
\Delta\Omega &= - \Tr\ln \bigg(
\uu_{cc} - \tilde\Sigma_{cc} \gf_{cc}
\bigg)\;.
\end{align*}

\section{\label{sec:AppendixFSR}Behavior of the Friedel sum rule within ED, CPT and VCA}
It is possible to  gain a somewhat deeper understanding of the behavior of the FSR within ED/CPT/VCA by considering the local Green's function at the impurity f-orbital
\begin{align}
 \GF_{ff}(z) &= \left( z - \epsilon_f -\Gamma(z)-\Sigma(z)\right)^{-1}\;\mbox{,}
\label{eq:GFSRstart}
\end{align}
where $\Gamma(z)$ is the contribution due to the single-particle terms of the hybridization and $\Sigma(z)$ the self-energy due to the local interaction (for details see \app~\ref{sec:AppendixTwoSiteAnalytical}). In the following we consider the particle-hole symmetric case. We are interested in the behavior of the retarded Green's function $\GF_{ff}^{\text{ret}}(\omega) = \GF_{ff}(\omega+i0^+)$ at the Fermi energy ($\omega=0$), which we investigate by taking the limit on the Matsubara axis $\GF_{ff}^{\text{ret}}(0) = \lim\limits_{\nu\rightarrow 0^+}\,\GF_{ff}(i\nu)$. We expand the self-energy $\Sigma(i\nu)$ up to linear order in $\nu$ and rewrite the expression using the definition of the effective quasi particle mass $m^*$ (see \eq{eq:meff})
\begin{align*}
 \Sigma(i\nu) &\approx -\epsilon_f  + i \Im{\text{m}(\Sigma(0))} + i\nu \frac{\partial \Im{\text{m}(\Sigma(i\nu))}}{\partial (i\nu)}\bigg|_{0^+} + i\mathcal{O}\left((i\nu)^2\right)\\
  &\approx -\epsilon_f + i \nu (1-m^*)\;\mbox{.}
\end{align*}
Inserting into \eq{eq:GFSRstart} we obtain
\begin{align}
 \nonumber\GF_{ff}^{\text{ret}}(0) &= \lim\limits_{\nu\rightarrow 0^+}\,\GF_{ff}(i\nu)\\
&=\lim\limits_{\nu\rightarrow 0^+}\,\Bigg( i \left[ m^*\nu -\Im{\text{m}(\Gamma(i\nu)} \right] - \Re{\text{e}(\Gamma(i\nu)} \Bigg)^{-1}\;\mbox{.}
\label{eq:GFSR2}
\end{align}
We will now investigate two separate, general cases of an ED and a CPT/VCA treatment of the Green's function.\\
From \eq{eq:GFSR2} it is easy to see that the remnant $m^*$ of the impurity self-energy and therefore the self-energy itself does not contribute to the Friedel sum rule. In outlining how to notice this, we simultaneously show that a discrete spectrum of the conduction band (as obtained for example in ED) will not fulfill the Fridel sum rule. Consider an arbitrary discrete spectrum of the conduction electrons with hybridization
\begin{align*}
 \Gamma(i\nu) &= V^2\,\sum\limits_\mu \frac{\alpha_\mu}{i\nu-\omega_\mu}\;\mbox{,}
\end{align*}
with excitation energies $\omega_\mu$. Splitting into real and imaginary parts and inserting into \eq{eq:GFSR2} gives
\begin{align*}
 \Im{\text{m}(\GF_{ff}^{\text{ret}}(0))} &= \lim\limits_{\nu\rightarrow 0^+}\\
&\frac{-\left(m^*+V^2 A(\nu)\right)\nu}{\left(m^{*2}+2m^*V^2A(\nu)+V^4A(\nu)^2\right)\nu^2+V^4B(\nu)^2}\\
\mbox{with}\\
A(\nu) &= \sum\limits_\mu \alpha_\mu \frac{1}{\omega_\mu^2+\nu^2}\;\mbox{, and}\\
B(\nu) &= \sum\limits_\mu \alpha_\mu \frac{\omega_\mu}{\omega_\mu^2+\nu^2}\;\mbox{.}
\end{align*}
Upon neglecting the weak dependence of $A(\nu)$ and $B(\nu)$ on $\nu$ one obtains, in this case, in the limit $\nu \rightarrow 0$: $\Im{\text{m}(\GF_{ff}^{\text{ret}}(0))}\rightarrow 0$ if all $\omega_\mu\neq0$ or $\Im{\text{m}(\GF_{ff}^{\text{ret}}(0))}\rightarrow -\infty$ if any $\omega_\mu=0$. We would like to further illustrate this for the specific model considered in this work (i.e. an impurity coupled to a semi-infinite chain with open boundary conditions)
\begin{align*}
 \mathcal{H} &= -V ( c_f^\dag c_0^\nag + c_0^\dag c_f^\nag) -t\,\sum\limits_{i=0}^{L-1} ( c_i^\dag c_{i+1}^\nag + c_{i+1}^\dag c_i^\nag)\;\mbox{,}
\end{align*}
where we supressed spin indices. We obtain by using the equation of motion for the local impurity f-orbital Green's function
\begin{align*}
 \omega\,\GF_{ff}(\omega) &= \omega\,<<c_f^\nag;c_f^\dag>> \\
&= \langle [c_f^\nag,c_f^\dag]_+\rangle - <<[c_f^\nag,\mathcal{H}]_-;c_f^\dag>>\;\mbox{,}
\end{align*}
for an even number of sites (including the impurity)
\begin{align*}
 \Im{\text{m}(\Gamma_{\text{even}}(i\nu))} &= -\frac{V^2}{\nu}\, R_L(\nu,t)\;\mbox{,}
\end{align*}
and for an odd number of sites
\begin{align*}
 \Im{\text{m}(\Gamma_{\text{odd}}(i\nu))} &= -V^2\nu\, R_L(\nu,t)\;\mbox{,}
\end{align*}
where $R_L(\nu,t)$ is a rational function which is well behaved upon taking the limit $\nu\rightarrow 0^+$ (i.e. it approaches a constant $\lim\limits_{\nu\rightarrow 0^+}\,R_L(\nu,t) = f_L(t)$) and the real part is always zero. Upon inserting into \eq{eq:GFSR2} one can easily verify that for an even number of sites this yields zero spectral weight at the Fermi energy, while for an odd number of sites it yields $-\infty$ (i.e. there is a pole exactly at $\omega=0$). Therefore the ED results alternate with even/odd system size between $\rho_f^{\text{even}}(0)=0$ and $\rho_f^{\text{odd}}(0)=\infty$. This result shows that the FSR is always violated in ED because a finite value of the impurity density of states at the Fermi energy may only be obtained using an artificial numerical broadening. It furthermore shows that all terms involving $m^*$ go to zero and cannot contribute to the sum rule.\\
Now we turn to the case of CPT/VCA, where the conduction electron hybridization takes the form (see \eq{eq:env})
\begin{align*}
 \Gamma(i\nu) &= i\,\frac{V^2}{2t^2}\;\left(\nu-\sqrt{4t^2+\nu^2}\right)\mbox{,}
\end{align*}
for the model considered in \eq{eq:HSIAM} (i.e. a semi-circular density of states of the conduction electrons). The CPT/VCA Green's function of the physical system is then given upon insertion of this $\Gamma$ into \eq{eq:GFSR2}
\begin{align*}
 \Im{\text{m}(\GF_{ff}^{\text{ret}}(0))} &= \lim\limits_{\nu\rightarrow 0^+}\\
&\frac{-1}{\left(m^*-\frac{V^2}{2t^2}\right)\nu+\frac{V^2}{2t^2}\sqrt{4t^2+\nu^2}}\;\mbox{.}
\end{align*}
Which yields upon expansion of the square root up to linear order in $\nu$
\begin{align*}
 \Im{\text{m}(\GF_{ff}^{\text{ret}}(0))} &= \lim\limits_{\nu\rightarrow 0^+}\\
&\frac{-1}{\left(m^*-\frac{V^2}{2t^2}+\frac{1}{4t}\right)\nu+\frac{V^2}{t}}\;\mbox{,}
\end{align*}
and when the limit $\nu\rightarrow 0$ is taken
\begin{align*}
 \Im{\text{m}(\GF_{ff}^{\text{ret}}(0))} &= -\frac{t}{V^2} = -\frac{1}{\Delta}\;\mbox{,}
\end{align*}
which is exactly the value predicted by the FSR (\eq{eq:FSR})
\begin{align*}
 \Im{\text{m}(\GF_{ff}^{\text{ret}}(0))} &= -\pi\frac{1}{\pi\Delta}\sin{(\frac{\pi}{2})}^2 = -\frac{1}{\Delta}\;\mbox{.}
\end{align*}
This result is independent of the size of the reference system. Away from particle-hole symmetriy (occupation of the f-orbital not one) the calculation becomes more tedious. The numerical VCA calculations however show that a pinning of the Kondo resonance at the Fermi energy is obtained in contrast to CPT at small $L$.\\

\section{\label{sec:AppendixTwoSiteAnalytical}Semi-analytical expressions for VCA of the two-site problem}
\begin{figure}
        \centering
        \includegraphics[width=0.48\textwidth]{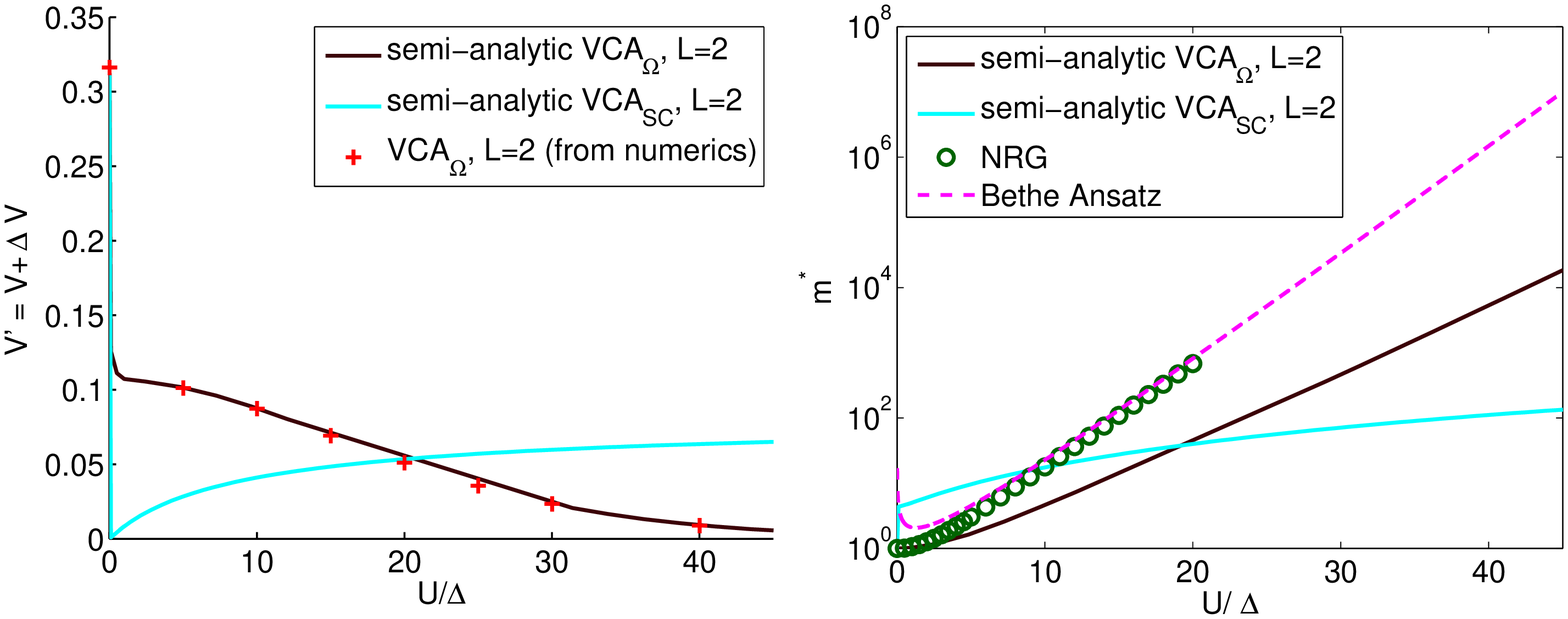}
        \caption{(Color online) Left: Optimal parameter $V'$ of the reference system as obtained by the semi-analytical equations for \vcaom \eq{eq:TwoSiteOmega} and \vcasc \eq{eq:TwoSiteSC}. As a reference the $L=2$ data of our numerical simulation is shown too. Right: The effective mass \eq{eq:effMassTwoSite1} obtained by the optimized parameter $V'$ of the reference system (see left figure). Additionally shown is the Bethe-Ansatz~\cite{hewson_kondo_1997} \eq{eq:Tk} and NRG~\cite{karrasch_finite-frequency_2008} result as a reference.}
        \label{fig:TwoSite}
\end{figure}

To gain a better understanding of the behavior of the low-energy properties of the SIAM we solve a small system semi-analytically. We obtain the scaling of the Kondo temperature $T_K$ with interaction strength $U$ within \vcaom as well as \vcascNOBLANK. A reference system consisting of a two site cluster and an infinite environment is used.\\
The Kondo scale will be determined from the effective mass \eq{eq:meff} using the self energy of a two site cluster \eq{eq:SigmaTwoSite} which leads to
\begin{align}
m^*(U) &= 1 + \frac{1}{36}\left(\frac{U}{V'}\right)^2\;\mbox{.}
\label{eq:effMassTwoSite1}
\end{align}
The Kondo scale $T_K$ is inversely proportional to $m^*(U)$ \eq{eq:TkTwoSite}. Therefore within this approximation the behavior of the optimal cluster parameter $V'$ governs the low energy physics.\\
To determine the optimal hopping $V'$ the Green's function of the reference system $\gf$ is calculated
\begin{align*}
\gf^{-1}(z) &= 
\begin{pmatrix}
z-\Sigma'(z)& V'& 0\\
V'& z& 0\\
0& 0& \GF'^{-1}_{ee}(z)\\
\end{pmatrix} \,\mbox{.}
\end{align*}
The CPT/VCA Green's function $\GF$ 
\begin{align*}
\GF^{-1}(z)& = 
\begin{pmatrix}
z-\Sigma'(z)& V& 0\\
V& z& t\\
0& t& \GF'^{-1}_{ee}(z)\\
\end{pmatrix} \,\mbox{, }
\end{align*}
is obtained by \eq{eq:Dyson} using 
\begin{align*}
\TF &=
\begin{pmatrix}
0 & \Delta V & 0\\
\Delta V & 0 & -t\\
0 & -t& 0\\
\end{pmatrix} \,\mbox{.}
\end{align*}
Here $\GF_{11}=\GF_{ff}$ \eq{eq:GffTwoSite} corresponds to the impurity f-orbital, $\GF_{22}=\GF_{ss}$ the second site in the interacting part of the reference system and $\GF_{33}=\GF_{ee}$ \eq{eq:env} the semi-infinite environment. The cluster parameter $V'$ is given by the physical parameter $V$ plus the variation $\Delta V$. Note that in here we work with the reduced expressions for $\Omega$ and $\GF$ justified in \app~\ref{sec:AppendixOmeganew}. Sch\"onhammer and Brenig calculated the Green's function of the correlated orbital for this model perturbativeley and showed that their expression becomes exact in the limit of vanishing bandwidth~\cite{brenig_theory_1974}. This is exactly the case considered here, where the impurity f-orbital is coupled to a single non interacting site providing a bath with vanishing bandwidth. They obtained 
\begin{align}
\gf_{ff}(z)=\frac{1}{z-\Gamma'(z)-\Sigma'(z)}\;\mbox{,}
\label{eq:GffTwoSite}
\end{align}
where the hybridization $\Gamma'(z)$ in our case is given by
\begin{align*}
\Gamma'(z) = \frac{V'^2}{z}\;\mbox{,}
\end{align*}
and the self-energy $\Sigma'(z)$ is given by
\begin{align}
\Sigma'(z) = \frac{\frac{U^2}{4}}{z-9\Gamma'(z)}\;\mbox{.}
\label{eq:SigmaTwoSite}
\end{align}
From this all elements of the cluster Green's function may be obtained by the equation of motion technique.\\
To be able to calculate the grand potential $\Omega$, the ground state energy of the interacting part of the reference system, $\omega'_0$ needs to be obtained (which can be done for example by diagonalization of the Hamiltonian matrix or by an integral over the Green's function)
\begin{align*}
\omega'_0&=-\frac{1}{4} \left(U + \sqrt{U^2 + 64 V'^2}\right)\;\mbox{.}
\end{align*}
Using the expression for the CPT/VCA Green's function $\gf$ and the ground state energy and taking the derivative of the grand potential \eq{eq:omegaResult} with respect to $\Delta V$ one is able to obtain an integral expression which allows to determine $V'$ within \vcaom:
\begin{align}
\nonumber \frac{d\,\Omega(\Delta V)}{d\,(\Delta V)} &= \nabla_{\Delta V}\omega_0'(\Delta V) -\frac{1}{\pi}\,\int_0^\infty\,d\omega\, \\
\nonumber &\Re{\text{e}}\Bigg(\tr\Bigg(\left(\uu-\TF(\Delta V)\gf(i\omega,\Delta V)\right)^{-1}\Bigg(\\
\nonumber &\left(\nabla_{\Delta V}\TF(\Delta V)\gf(i\omega,\Delta V)\right)\\
&+\left(\TF(\Delta V)\nabla_{\Delta V}\gf(i\omega,\Delta V)\right)\Bigg)\Bigg)\Bigg)\stackrel{!}{=}0\;\mbox{.}
\label{eq:TwoSiteOmega}
\end{align}
The resulting $V'(U)$ is shown in \fig{fig:TwoSite} (left) and is used to calculate the effective mass \eq{eq:effMassTwoSite1} shown right in the figure. The effective mass shows exponential behavior but the exponent does not match the Bethe Ansatz result as discussed in \se~\ref{sec:LowEnergy}.\\
Next we attempt to obtain the \vcasc solution for the two-site problem. The only variational parameter is $V$ and therefore we determine the expectation value of $\sum\limits_\sigma <f^\dag_\sigma c^\nag_{1\sigma}>$ self consistently. Here $1$ denotes the impurity's s-orbital. Since we are considering a spin-symmetric model we sum over both spin directions and denote this expectation value as $<f^\dag c>$ in the following. The hopping expectation value is given by
\begin{align*}
<f^\dag c> = -\frac{2}{\pi}\,\int_0^\infty\,d\omega\,\GF_{fc}(i\omega)\;\mbox{.}
\end{align*}
Evaluation of this expectation value in the cluster yields
\begin{align*}
<f^\dag c>_{\text{cluster}} = -\frac{2}{\pi}\,\int_0^\infty\,d\omega\,\\
\frac{4 (\Delta V+V) \left(9 (\Delta V+V)^2+w^2\right)}{36 (\Delta V+V)^4+\left(U^2+40 (\Delta V+V)^2\right) w^2+4 w^4}\;\mbox{.}
\end{align*}
Evaluation of this expectation value in the total system gives
\begin{align*}
<f^\dag c>_{\text{CPT}} &= -\frac{2}{\pi}\,\int_0^\infty\,d\omega\\
&\frac{8 V}{8 V^2+\frac{w \left(U^2+36 (\Delta V+V)^2+4 w^2\right) \left(w+\sqrt{4 t^2+w^2}\right)}{9 (\Delta V+V)^2+w^2}}\;\mbox{.}
\end{align*}
Upon requiring the two expectation values to coincide
\begin{align}
<f^\dag c>_{\text{cluster}} \stackrel{!}{=} <f^\dag c>_{\text{CPT}}\;\mbox{,}
\label{eq:TwoSiteSC}
\end{align}
the optimal value of $\Delta V$ is obtained numerically. The resulting $V'(U)$ is plotted in \fig{fig:TwoSite} (left) and is used to calculate the effective mass \eq{eq:effMassTwoSite1} shown right in the figure. The effective mass does not show exponential behavior in \vcascNOBLANK.
 

\end{document}